\documentclass[a4paper,11pt]{article}
\pdfoutput=1 

\usepackage{jcappub} 

\usepackage[T1]{fontenc} 

\usepackage{aas_macros} 
\usepackage{lineno}

\usepackage[utf8]{inputenc}
\usepackage{longtable}
\usepackage{rotating}
\usepackage{float} 
\usepackage{graphicx}
\usepackage{subcaption} 
\usepackage{placeins}   

\usepackage{booktabs}
\usepackage{rotating}   
\usepackage{adjustbox}
\usepackage{accents}

\title{Gamma-Ray Constraints on Heavy Axion-Like-Particle Decays from \textit{Fermi}-LAT and H.E.S.S. Blazar Spectra}

\author[a,1]{A.~Acharyya}
\author[c,d,e]{F.~Aharonian}
\author[f,g]{M.~Backes}
\author[h]{R.~Batzofin}
\author[i]{Y.~Becherini}
\author[j]{S.~Bisero}
\author[g]{M.~B\"ottcher}
\author[k]{C.~Boisson}
\author[l]{J.~Bolmont}
\author[j]{F.~Brun}
% ...

\affiliation[a]{CP3-Origins, University of Southern Denmark, Campusvej 55, 5230 Odense M, Denmark}
\affiliation[c]{Yerevan State University, 1 Alek Manukyan St, Yerevan 0025, Armenia}
\affiliation[d]{Max-Planck-Institut f\"ur Kernphysik, P.O. Box 103980, D 69029 Heidelberg, Germany}
\affiliation[e]{Dublin Institute for Advanced Studies, Dublin D15 XR2R, Ireland}
\affiliation[f]{University of Namibia, Department of Physics, Windhoek 10005, Namibia}
\affiliation[g]{Centre for Space Research, North-West University, Potchefstroom 2520, South Africa}
\affiliation[h]{Universit\"at Potsdam, D 14476 Potsdam, Germany}
\affiliation[i]{Université Paris Cité, CNRS, APC, F-75013 Paris, France}
\affiliation[j]{IRFU, CEA, Université Paris-Saclay, F-91191 Gif-sur-Yvette, France}
\affiliation[k]{LUX, Observatoire de Paris, 92190 Meudon, France}
\affiliation[l]{Sorbonne Université, CNRS/IN2P3, LPNHE, 75005 Paris, France}
% ...

\emailAdd{contact.hess@hess-experiment.eu}

\note{Corresponding authors.}

\abstract{
The propagation of very-high-energy (VHE; $E_{\gamma} \geq 100$~GeV) gamma rays from extragalactic sources is affected by interactions with photons of the extragalactic background light (EBL), resulting in pair production that attenuates the intrinsic gamma-ray flux. This interaction renders the Universe increasingly opaque to VHE photons at high energies and redshifts. New physics scenarios involving axion-like particles (ALPs) could modify this expected optical depth. In particular, ALPs with masses $m_a \sim 10$~eV can decay into two photons over cosmological timescales, thereby contributing to the diffuse EBL. If such ALPs constitute a significant fraction of the dark matter density, their decay would enhance the EBL intensity and consequently increase the gamma-ray optical depth. In this study, we investigate this scenario using a large sample of gamma-ray spectra observed with the High Energy Stereoscopic System (H.E.S.S.) and the \textit{Fermi} Large Area Telescope. We model the contribution of decaying ALPs to the EBL and assess their impact on the spectra of blazars across redshifts. By comparing these observations with standard EBL models, we place constraints on the properties of heavy ALPs—specifically their mass and photon coupling—and evaluate their viability as a dark matter candidate capable of modifying the gamma-ray transparency of the Universe. From the combined analysis, and under the assumption that ALPs constitute the entire dark matter density, we derive 95\%~confidence exclusion limits on the photon--ALP coupling down to $g_{a\gamma}\sim7\times10^{-12}$~GeV$^{-1}$ for masses $m_a\sim15$~eV. 
These constraints are competitive with existing astrophysical bounds and provide complementary sensitivity to other techniques, closing a previously unconstrained region of parameter space in the $m_a \sim 2.5$--$20$~eV range.
}


\usepackage{xcolor} 
\usepackage{graphicx}
\usepackage[dvipsnames]{xcolor}

\usepackage[numbers]{natbib}
\usepackage{hyperref}

\usepackage{booktabs}

\usepackage{txfonts}

\begin{document}
\linenumbers
\maketitle
\flushbottom

\section{Introduction}

Very-high-energy (VHE; $E_{\gamma} \geq 100$~GeV) gamma rays emitted by extragalactic sources are expected to undergo attenuation as they traverse intergalactic space. This attenuation arises from interactions with the diffuse photon field known as the extragalactic background light (EBL), which spans ultraviolet to far-infrared wavelengths. The dominant process---electron--positron pair production---was studied in the astrophysical context in pioneering works such as \cite{Nikishov1962,Gould1966,Gould1967}. This process leads to an energy- and redshift-dependent suppression of the observed gamma-ray flux, imposing a fundamental limitation on our ability to observe distant, highly energetic sources. For a review of the EBL and its impact on gamma-ray propagation, see \cite{Dwek2013}.

Accurately modelling this EBL-induced attenuation is thus critical in gamma-ray astronomy---not only to infer the intrinsic spectra of extragalactic sources such as blazars, but also to test our understanding of photon propagation over cosmological distances (see, e.g., \cite{Biteau2022} for a recent review). 
Direct measurement of the EBL through photometry is severely hampered by strong foreground contamination, particularly from zodiacal light \cite{Hauser1998}.
Consequently, several EBL models based on galaxy evolution, spectral energy distributions (SEDs), and deep surveys have been developed and widely adopted (e.g., \cite{Franceschini_2008, 2010ApJ...712..238F, 2011MNRAS.410.2556D, 2021MNRAS.507.5144S}). The attenuation of VHE gamma rays by the EBL has also been used to constrain the EBL itself using blazar spectra. In particular, the High Energy Stereoscopic System (H.E.S.S.) Collaboration performed a likelihood-based measurement of the EBL SED by extracting the characteristic energy- and redshift-dependent absorption imprint from high-quality VHE blazar spectra \cite{2017A&A...606A..59H}.

Despite the success of these models, potential anomalies have been discussed (see, e.g., \cite{Biteau2022}), with several individual gamma rays at energies where optical depth is high, appearing inconsistent with the expected EBL attenuation. For example, the unexpected TeV detection of GRB~221009A by LHAASO and PeV-scale events from the Cygnus region \cite{2023Sci...380.1390L, 2023EPJWC.28001003C}, were discussed in the literature. These discussions have spurred theoretical interest in beyond-standard-model physics that could modify the gamma-ray transparency of the Universe. It should be noted that recent measurements of the cosmic optical background by the instrument on the New Horizons mission indicate that the previously reported optical EBL intensity excess is largely resolved once improved modelling of diffuse Galactic light and faint-source contributions is applied \cite{2024ApJ...972...95P}.

Nevertheless, the level and spectral shape of the EBL provides a sensitive probe for new physics, and one compelling possibility involves axion-like particles (ALPs)---hypothetical pseudoscalar bosons predicted in various extensions of the Standard Model. 
In the presence of astrophysical magnetic fields, photons may oscillate into ALPs and vice versa \cite{Raffelt1988}.
These photon--ALP conversions can allow gamma rays to bypass EBL attenuation over long distances, only to reconvert to photons closer to Earth, effectively increasing the Universe's transparency. 
This mechanism has been widely explored in the literature, from phenomenological studies of photon–ALP mixing and its impact on gamma-ray propagation \cite{Galanti_2018,Roncadelli2009,Hooper2007,Mirizzi2007} to observational hints of spectral hardening in distant VHE blazars \cite{Meyer_2014,2009PhRvD..79l3511S,2011JCAP...11..020D}, as well as searches for anomalously transparent gamma-ray propagation using the highest-energy photons detected by space-based gamma-ray telescopes (for example \cite{2020JCAP...09..027B}).

Their interaction with photons is described by the effective Lagrangian
\begin{equation}
\label{eq:LaGamma}
\mathcal{L}_{a\gamma} = -\tfrac{1}{4}\, g_{a\gamma}\, a\,F_{\mu\nu}\,\tilde{F}^{\mu\nu}
= g_{a\gamma}\,a\,\boldsymbol{E}\!\cdot\!\boldsymbol{B}\,,
\end{equation}
where $a$ is the ALP field, $F_{\mu\nu}$ the electromagnetic field--strength tensor, and
$\tilde F^{\mu\nu} = \tfrac{1}{2}\,\epsilon^{\mu\nu\rho\sigma} F_{\rho\sigma}$ its dual.
In the last equality, $\boldsymbol{E}$ and $\boldsymbol{B}$ denote the electric and magnetic fields,
such that $F_{\mu\nu}\,\tilde F^{\mu\nu} \,=\, -4\,\boldsymbol{E}\!\cdot\!\boldsymbol{B}$ for our metric convention.
Here, $g_{a\gamma}$ is the photon--ALP coupling constant.
Following common practice in the literature, we denote the ALP--photon coupling as
\(g_{a\gamma}\); this corresponds to the two--photon coupling \(g_{a\gamma\gamma}\) used in some works.

In this paper, we explore a complementary but comparatively less-studied scenario enabled by
the interaction in Eq.~\eqref{eq:LaGamma}: the decay of heavy ALPs (with masses
$m_a \sim 10$~eV) into two photons.
This operator mediates the two--photon decay $a \to \gamma\gamma$, with a decay width that depends
on the ALP mass $m_a$ and the photon coupling $g_{a\gamma}$ (see Section \ref{alps_decay_theory} for details).
Such decays can inject additional photons into the cosmic infrared background
(see, e.g., \cite{Bernal2023}), enhancing the EBL intensity and leading to increased
attenuation of VHE gamma rays from distant sources.
Although photon--ALP mixing and ALP decay both arise from the same coupling term, in the heavy-ALP mass range considered here ($m_a \sim 2.5$--$20~\mathrm{eV}$) the mixing effect is strongly suppressed in astrophysical magnetic fields at GeV--TeV energies due to the large ALP mass term. Observable mixing signatures, such as spectral irregularities or upturns, are instead expected for much lighter ALPs in the neV mass range. We therefore neglect photon--ALP mixing and focus exclusively on the decay-induced contribution to the EBL.

We search for evidence of this decay-induced optical depth by analysing the gamma-ray SEDs of blazars. 
Our approach is to model the EBL including the contribution from ALP decays and compare it to standard models, testing whether the data show any statistical preference for this hypothesis. 
Furthermore, we scan the ALP mass--coupling parameter space, allowing us to place constraints on the properties of heavy ALPs and assess their viability as a dark--matter component influencing the propagation of VHE gamma rays across the Universe.

To perform this search, we use data from two complementary gamma-ray instruments. 
The \emph{Fermi}--Large Area Telescope (LAT; \cite{Fermi_LAT}) is a pair-conversion gamma-ray detector sensitive to photons from 20~MeV to beyond 500~GeV. Operating primarily in survey mode, it scans the entire sky every three hours, providing uniform exposure and long-term monitoring of extragalactic sources. 
At higher energies, we rely on the High Energy Stereoscopic System (H.E.S.S.), an array of five imaging atmospheric Cherenkov telescopes (IACTs) located in Namibia \citep{2023hxga.book..142P}. 
The system consists of four 12~m telescopes in a 120~m square and a central 28~m telescope (CT5) that lowers the energy threshold. 
Together, H.E.S.S.\ provides sensitivity from $\sim$30~GeV to 100~TeV, with $\sim$15\% energy resolution and angular resolution better than $0.1^{\circ}$ \citep{2014APh....56...26P}. 
This work represents the first analysis of heavy ALP decay using both \emph{Fermi}--LAT and H.E.S.S.\ data, combining both instruments into a unified dataset and maximising the joint likelihood directly.
This joint approach allows us to maximise statistical robustness across a wide energy range, providing robust constraints on heavy ALP models that may alter the gamma-ray optical depth of the Universe.

\section{Data: Selection, Analysis, and Segmentation}
\label{data_analysis_pipeline}

\subsection{Source Selection}

The identification of suitable sources for any ALP- or EBL-related study is primarily governed by the need for sufficient photon statistics to enable robust investigation of gamma-ray spectra. In particular, we prioritise targets at distances where absorption becomes significant, thereby requiring sources at very high energies and redshifts.

We constructed our blazar sample from targets observed with H.E.S.S.\ for which the highest-energy spectral data points fall within the optically thick regime, defined by optical depths $\tau>2$. This criterion prioritised sensitivity to potential deviations from standard absorption expectations, which were expected to manifest most clearly at large $\tau$. While our initial list was based on published reports of VHE flaring activity during H.E.S.S.\ campaigns, our analysis is not restricted to flaring states. Instead, we consolidated all available observations in the vicinity of these episodes—both during and surrounding the reported high states—rather than restricting strictly to the flare windows. This provided adequate coverage at lower optical depths as well, which was essential to model the intrinsic source spectra consistently. To minimise sources of systematic errors, we further required sources to be detected at zenith angles $\theta_{z}<60^{\circ}$. 

The final sample consists of 11 sources, listed in Table~\ref{tab:combined_blazar_summary}, along with their equatorial coordinates and cosmological redshifts, primarily obtained from the Simbad database \cite{2000A&AS..143....9W}.
It is important to emphasise that the goal of this study is not to characterise the behaviour of individual blazars, but rather to examine how ALPs may affect gamma-ray transparency through their impact on the EBL optical depth.

\subsection{H.E.S.S. Observations and Analysis}
\label{data_analysis_pipeline_2}

Following \cite{Aharonian:2023hessIGMF}, we analysed a homogeneous four-telescope (CT1–4) data set, explicitly excluding CT5 to avoid configuration-dependent systematics. A significant fraction of the analysed data were taken before the installation of CT5.
Only good-quality runs with at least three participating telescopes and zenith angles $\theta_{z} < 60^{\circ}$ were retained. Our data and source selection procedure followed \cite{Aharonian:2023hessIGMF}, but we additionally included sources observed in flaring states, and consolidated observations before and after these episodes to maximise statistics and enable stable intrinsic spectral modelling.

The energies of the recorded gamma-ray–like events were reconstructed using the \texttt{ImPACT} method with standard selection cuts \citep{2015ICRC...34..826P}. Event selection was based on a circular region of interest (ROI) of radius $0.071^{\circ}$ centred on the source position, corresponding to the 68\% containment radius of the H.E.S.S.\ point-spread function for point-source analyses. 
All observations were taken in wobble mode with typical offsets of $0.5^{\circ}$ from the source position, enabling simultaneous background estimation. The background was evaluated using the reflected-region method \citep{berge2007}, where multiple off-source regions at the same radial distance from the camera center as the on-source region provide a robust estimate of the residual cosmic-ray background.
In addition to statistical uncertainties, systematic effects primarily related to the H.E.S.S.\ energy scale and EBL modelling, are considered in this analysis. These are incorporated within the likelihood framework and discussed in more detail in Section~\ref{alps_decay_theory}.

The reduced event lists were converted into FITS format to ensure compatibility with the \texttt{Gammapy} package version 1.1 \citep{gammapy:2023, aguasca_cabot_2023_8033275}, which was then used to perform spectral modelling and temporal variability analysis. 
Nightly-binned light curves were constructed by fitting power-law spectra to compute differential fluxes. These light curves provided the basis for assessing both short-term flaring activity and long-term variability in the VHE band. They also served as the starting point for our temporal segmentation procedure (Section~\ref{segmentation}), where we identified statistically significant changes in flux and, where appropriate, investigated accompanying spectral evolution.
As a consistency test, the spectral and temporal results obtained with the HAP (Heidelberg) chain were cross-checked using the independent HAP-Fr (Paris) analysis framework \citep{Khelifi:2015pzi}, which were found to agree within statistical uncertainties.

\begin{figure}[!p]
    \centering
    \includegraphics[width=0.95 \linewidth]{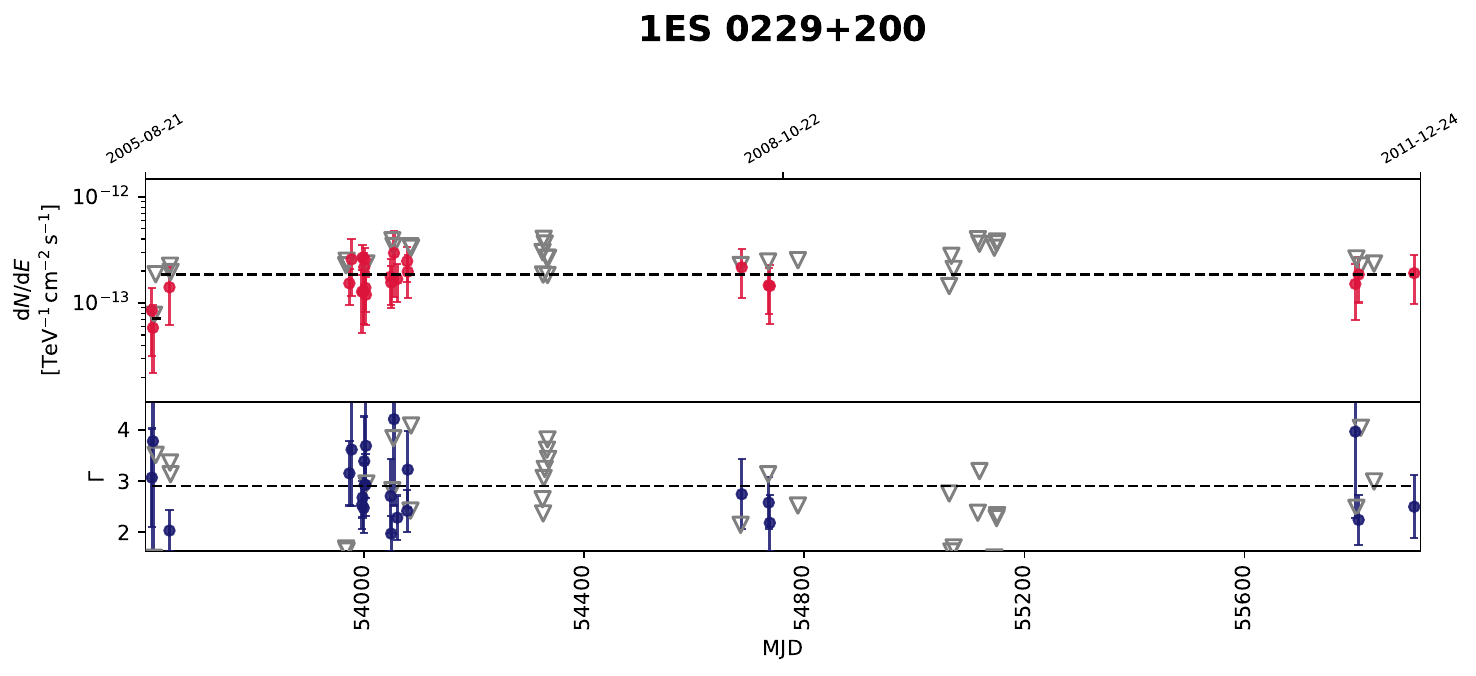}
    \includegraphics[width=0.95 \linewidth]{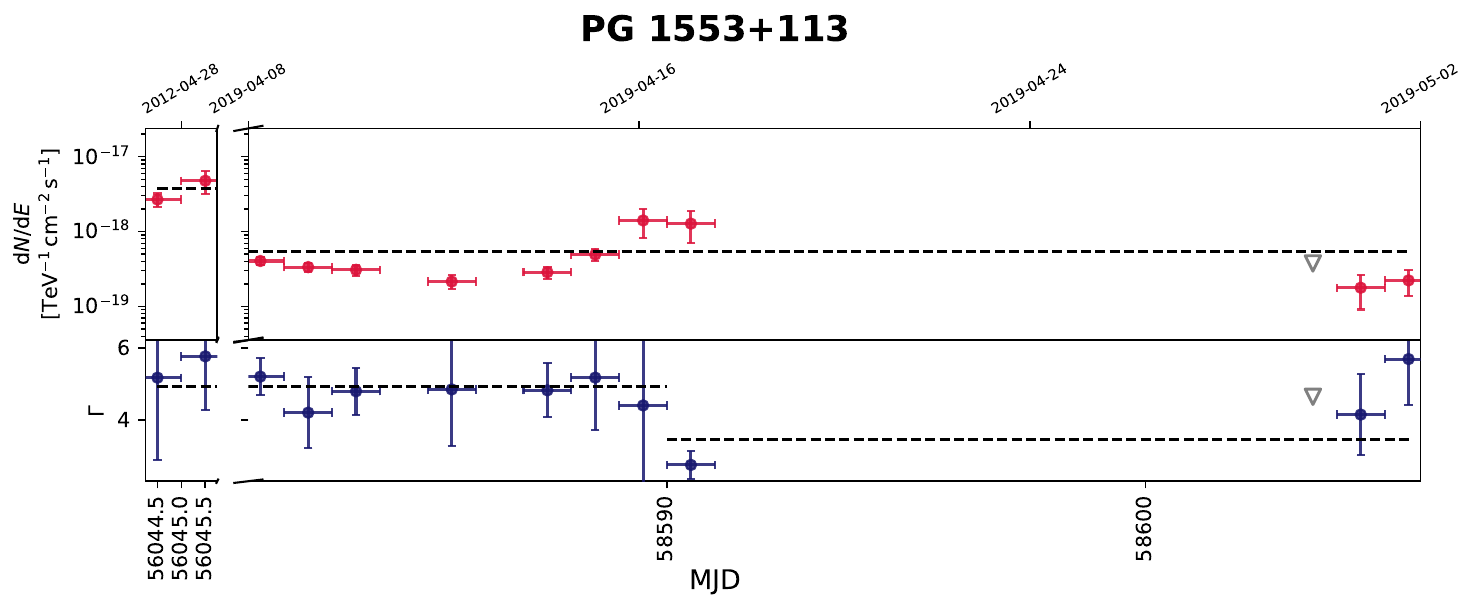}
    \includegraphics[width=0.95 \linewidth]{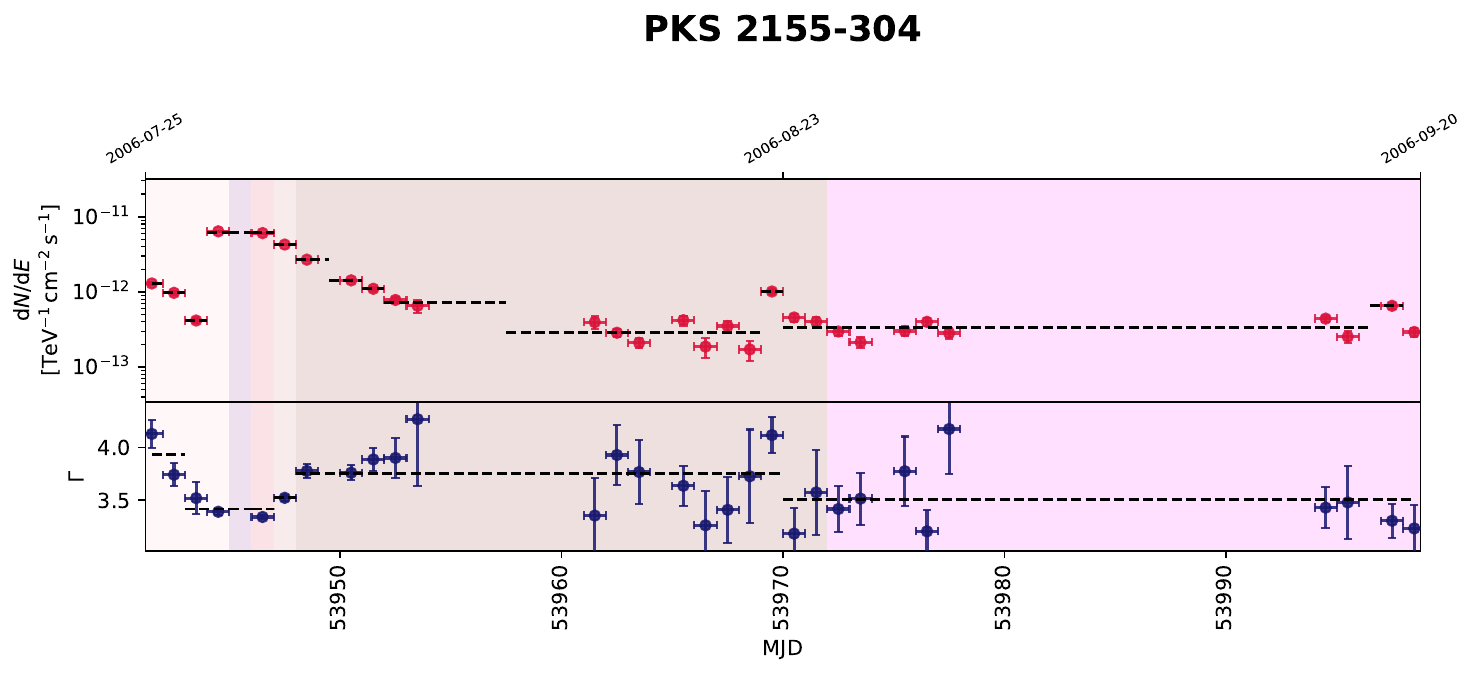}
    \caption{Representative examples of the temporal segmentation procedure applied to H.E.S.S.\ data.
    In all panels, the upper subplots show the nightly-binned flux evolution, with red points indicating detections
and grey inverted triangles denoting 95\% upper limits. Horizontal dashed lines indicate the Bayesian-block
segmentation of the flux. The lower subplots show the evolution of the photon index~$\Gamma$, with dashed lines
indicating the corresponding Bayesian-block segmentation. For PKS~2155--304, the shaded regions additionally indicate the six time intervals used in the spectral analysis and reported in Table~1.
\textbf{Top:} 1ES~0229+200, a relatively quiescent source.
\textbf{Middle:} PG~1553+113, exhibiting moderate flux variability.
\textbf{Bottom:} PKS~2155--304, a strongly variable source.}
    \label{fig:lc_representative}
\end{figure}

We also compared our reconstructed spectra with previously published H.E.S.S. results, including those from the H.E.S.S. Extragalactic Survey (HEGS; \citep{refId0}) and earlier analyses (listed in the final column of Table~\ref{tab:combined_blazar_summary}). Overall, we find good agreement within statistical uncertainties for the vast majority of sources, in particular for non-variable sources or comparable flux states. For variable sources, differences are expected due to the use of flare-selected intervals in this work compared to long-term averaged data in HEGS, while comparisons with analyses based on similar flux states also show good agreement. Minor differences are limited to a few cases, namely PG~1553+113 and PKS~1510$-$089, and can be understood in terms of non-contemporaneous data selections, as well as differences in instrument configurations (e.g.\ the inclusion of CT5 in earlier analyses).
Figure~\ref{fig:lc_representative} illustrates representative examples of the temporal segmentation procedure, showing a quiescent source, a moderately variable source, and a strongly variable source. The full set of light curves for all analysed sources is provided in Appendix~\ref{Appendix}.

\subsection{Resulting Segmentation}
\label{segmentation}

Having constructed nightly light curves for each source, we evaluated whether temporal segmentation was warranted based on statistically significant flux variations. To identify such changes, we applied the Bayesian blocks algorithm implemented in \texttt{astropy} v6.1 \citep{Scargle:2012bb}, which adaptively partitions each light curve into segments (“blocks”) characterised by distinct flux levels. For non-variable sources, the algorithm returns a single block, indicating a quiescent state. Other sources exhibit more complex behaviour: some showed modest variability with few transitions, while others displayed pronounced flux changes across multiple blocks.

While Bayesian blocks provide an objective, data-driven segmentation, they do not account for potential spectral evolution. For sources with two or more blocks, we performed a spectral analysis within each block to assess whether further segmentation was necessary to avoid spectral-averaging biases. Each block was fit with a simple power-law model using \texttt{Gammapy}, from which we extracted the photon index and flux normalisation. To visualise spectral trends, we plotted the photon index versus the corresponding flux normalisation for each block. This diagnostic helped assess whether flux changes were accompanied by spectral evolution.

We examined whether spectral parameters varied by more than $2\sigma$ between adjacent Bayesian blocks, based on independent fits to each block. The merging was performed sequentially: when two adjacent blocks were consistent within $2\sigma$, they were combined and the resulting interval was then compared to the next neighbouring block. Blocks that were not consistent within $2\sigma$ were retained as separate intervals. Each source was reviewed manually to determine whether adjacent blocks should be merged or treated separately. To illustrate this procedure, Figure~\ref{fig:bestfit_ellipses_pks2155} in Appendix~\ref{Appendix} shows the best-fit photon index $\Gamma$ versus flux normalisation $N_0$ for PKS\,2155$-$304, with error ellipses for different temporal blocks. This representation highlights how adjacent blocks are compared: blocks with overlapping ellipses are typically merged, while those separated by more than $2\sigma$ in either parameter motivate retaining separate blocks to capture spectral evolution.

\subsection{\emph{Fermi}--LAT Observations and Data Analysis}

To extend our spectral coverage to lower energies, we incorporated \emph{Fermi}--LAT observations alongside H.E.S.S.\ data. 
If a source was consistent with a single, non-variable segment in H.E.S.S.\ (as determined by the Bayesian blocks analysis), we used 14~years of time-averaged \emph{Fermi}--LAT data, covering the mission start on August~4,~2008 to August~4,~2022 (MJD~54683--MJD~59796). 
For variable sources with two or more segments, we instead used \emph{Fermi}--LAT data contemporaneous with each H.E.S.S.\ segment, in order to avoid mixing different spectral states. 
 For segments predating the \emph{Fermi}--LAT mission (e.g.\ in PKS~2155$-$304), only H.E.S.S.\ data were used.
All \emph{Fermi}--LAT data were analysed using \texttt{Gammapy}, with cross-validation against results obtained using the \emph{Fermi} Science Tools v2.2.0\footnote{\url{http://fermi.gsfc.nasa.gov/ssc/data/analysis/software} (accessed on March~21,~2025)} and \texttt{Fermipy} v1.2\footnote{\url{http://fermipy.readthedocs.io} (accessed on March~21,~2025)} \citep{wood2017fermipy}. 
We adopted the latest \emph{Pass}~8 instrument response functions (IRFs; \cite{atwood2013pass}).

We performed a binned maximum-likelihood analysis in the 1--300~GeV energy range, using photons within a $10^{\circ}$ ROI centred on each source. 
Events were restricted to zenith angles $\theta_z < 90^{\circ}$ to reduce contamination from gamma rays originating at the Earth's limb, produced by cosmic-ray interactions in the upper atmosphere. 
All sources from the 4FGL-DR4 catalogue \citep{4fgl_dr3} within $20^{\circ}$ of the ROI centre were included in the background model, initialised with their catalogue spectral parameters. 
This treatment accounted for gamma-ray emission from sources lying outside the ROI that may contribute photons to the data, particularly at low energies owing to the size of the \emph{Fermi}--LAT point-spread function. 
To identify any significant uncatalogued excesses, we applied the \texttt{gtfindsrc} routine. 
No significant unmodelled excesses were found in any field; consequently, no additional sources were added to the model.

Spectral parameters were initialised to their 4FGL-DR4 values and re-optimised in the global binned-likelihood fit. 
The target source and all 4FGL objects within the $10^{\circ}$ ROI had free normalisations and spectral indices, whereas sources between $10^{\circ}$ and $20^{\circ}$ were kept fixed at their catalogue values. 
This approach minimised cross-contamination from bright neighbouring sources while avoiding over-parameterisation of the outer field.
The diffuse backgrounds were modelled using the standard templates: \texttt{gll\_iem\_v07.fits} for Galactic diffuse emission and \texttt{iso\_P8R3\_SOURCE\_V3\_v1.txt} for the isotropic component. 
The normalisations of these diffuse components, along with the flux normalisations of all sources within the ROI, were left free during the fit. 
The data were spatially binned in $0.1^{\circ}$ pixels and energy-binned at two bins per decade.

\begin{table*}[t]
\centering
\begin{adjustbox}{max width=\textwidth}
\begin{tabular}{l l c c c c c c c l}
\toprule
\textbf{Source Name} &
\textbf{Observation Range} &
\textbf{Redshift} &
$E_{\tau=2}$ [TeV] &
$N_0$ [$10^{-8}$~cm$^{-2}$~s$^{-1}$~TeV$^{-1}$] &
$\alpha$ &
$\beta$ &
$E_0$ [GeV] &
$s_{\mathrm{H.E.S.S.}}$ &
\textbf{Ref.} \\
\midrule
1ES 0229+200 & 21.08.2005 -- 24.12.2011 & 0.14 & 1.70 & $2.92 \pm 0.45$ & $1.71 \pm 0.06$ & -- & 3.9 & $-0.091 \pm 0.109$ & \text{\cite{2007A&A...475L...9A}}  \\
1ES 0347-121 & 26.08.2006 -- 20.12.2009 & 0.19 & 0.90 & $3.19 \pm 0.33$ & $1.78 \pm 0.06$ & -- & 3.6 & $-0.161 \pm 0.071$  & \text{\cite{2007A&A...473L..25A}} \\
1ES 0414+009 & 01.01.2007 -- 27.11.2009 & 0.29 & 0.53 & $5.85 \pm 0.49$ & $2.01 \pm 0.06$ & -- & 3.5 & $-0.160 \pm 0.067$ & \text{\cite{2012A&A...538A.103H}} \\
1ES 1101-232 & 15.04.2004 -- 23.01.2008 & 0.19 & 0.90 & $1.84 \pm 0.16$ & $1.78 \pm 0.05$ & -- & 5.7 & $-0.144 \pm 0.073$ & \text{\cite{2007A&A...470..475A}} \\
1ES 1312-423 & 15.04.2004 -- 23.06.2010 & 0.11 & 3.49 & $1.43 \pm 0.14$ & $1.93 \pm 0.05$ & -- & 6.0 & $\phantom{-}0.067 \pm 0.101$ & \text{\cite{2013MNRAS.434.1889H}} \\
1RXS J101015.9-311909 & 06.04.2004 -- 23.06.2010 & 0.14 & 1.70 & $2.07 \pm 0.16$ & $1.89 \pm 0.04$ & -- & 5.9 & $\phantom{-}0.163 \pm 0.092$ & \text{\cite{2012A&A...542A..94H}} \\
H2356-309 & 16.06.2004 -- 26.09.2006 & 0.17 & 1.08 & $0.93 \pm 0.41$ & $2.19 \pm 0.15$ & -- & 474.8 & $\phantom{-}0.046 \pm 0.137$ & \text{\cite{2010A&A...516A..56H}} \\
H2356-309 & 14.07.2007 -- 27.07.2007 & 0.17 & 1.08 & $0.87 \pm 0.39$ & $1.65 \pm 0.16$ & -- & 474.8 & $\phantom{-}0.023 \pm 0.162$ & \text{\cite{2010A&A...516A..56H}} \\
PG 1553+113 & 26.04.2012 -- 28.04.2012 & 0.43 & 0.36 & $3.28 \pm 0.21$ & $1.40 \pm 0.07$ & $0.08 \pm 0.02$ & 1.8 & $-0.163 \pm 0.062$ & \text{\cite{2006A&A...448L..19A}} \\
PG 1553+113 & 08.04.2019 -- 02.05.2019 & 0.43 & 0.36 & $4.83 \pm 0.23$ & $1.59 \pm 0.05$ & $0.03 \pm 0.01$ & 1.8 & $\phantom{-}0.010 \pm 0.046$ & \text{\cite{2006A&A...448L..19A}}  \\
PKS 0447-439 & 09.11.2009 -- 07.12.2010 & 0.34 & 0.46 & $3.32 \pm 0.16$ & $1.99 \pm 0.06$ & -- & 1.6 & $-0.128 \pm 0.065$  & \text{\cite{2013A&A...552A.118H}} \\
PKS 1510-089 & 06.04.2016 -- 27.07.2016 & 0.36 & 0.43 & $39.4 \pm 2.2$ & $2.51 \pm 0.04$ & -- & 0.9 & $-0.233 \pm 0.052$ & \text{\cite{2021A&A...648A..23H}} \\
PKS 2155-304 & 07.07.2006 -- 28.07.2006 & 0.12 & 2.78 & $(1.1 \pm 0.2)\times10^{3}$ & $2.49 \pm 0.06$ & $0.05 \pm 0.01$ & 25.8 & $\phantom{-}0.122 \pm 0.047$ & \text{\cite{2013PhRvD..88j2003A}} \\
PKS 2155-304 & 29.07.2006 -- 29.07.2006 & 0.12 & 2.78 & $(3.6 \pm 0.6)\times10^{3}$ & $2.04 \pm 0.10$ & $0.66 \pm 0.04$ & 572.2 & $-0.040 \pm 0.047$ & \text{\cite{2012A&A...539A.149H}} \\
PKS 2155-304 & 30.07.2006 -- 30.07.2006 & 0.12 & 2.78 & $(7.3 \pm 1.1)\times10^{3}$ & $2.24 \pm 0.09$ & $0.18 \pm 0.02$ & 93.0 & $\phantom{-}0.087 \pm 0.045$ & \text{\cite{2013PhRvD..88j2003A}} \\
PKS 2155-304 & 31.07.2006 -- 31.07.2006 & 0.12 & 2.78 & $1.14 \pm 0.35$ & $2.70 \pm 0.09$ & $0.33 \pm 0.07$ & 477.1 & $-0.123 \pm 0.063$ & \text{\cite{2013PhRvD..88j2003A}} \\
PKS 2155-304 & 01.08.2006 -- 24.08.2006 & 0.12 & 2.78 & $2.37 \pm 0.48$ & $2.81 \pm 0.09$ & $0.17 \pm 0.06$ & 222.1 & $-0.250 \pm 0.011$ & \text{\cite{2012A&A...539A.149H}} \\
PKS 2155-304 & 25.08.2006 -- 20.09.2006 & 0.12 & 2.78 & $1.66 \pm 0.76$ & $2.41 \pm 0.25$ & $0.11 \pm 0.12$ & 219.5 & $-0.048 \pm 0.103$ & \text{\cite{2010A&A...520A..83H}} \\
\bottomrule
\end{tabular}
\end{adjustbox}
\caption{
Summary of the blazar sample analysed in this work.
For each source, we list the redshift, the observing intervals corresponding to the merged Bayesian blocks, and for each period the best-fit spectral parameters obtained under the no-ALP hypothesis from a joint \textit{Fermi}--LAT and H.E.S.S.\ analysis where available, and from H.E.S.S.\ data alone otherwise. We also report $E_{\tau=2}$, defined as the energy at which the gamma-ray optical depth reaches $\tau_{\gamma\gamma}=2$ for the EBL model in \cite{2011MNRAS.410.2556D}.
For PKS 2155$-$304, some intervals correspond to individual nights identified through the temporal segmentation procedure.
Each spectrum was fitted with both a power-law and a log-parabola model, and the log-parabola model was adopted only when it provided a statistically significant improvement over the power-law fit, corresponding to $3\sigma$ significance based on the likelihood-ratio test.
The parameters $N_0$, $\alpha$, $\beta$, and $E_0$ denote the spectral normalisation (in units of $10^{-8}$~cm$^{-2}$~s$^{-1}$~TeV$^{-1}$), photon index, curvature parameter, and pivot energy, respectively (see Eqn \ref{eq:powerlaw} and Eqn \ref{eq:logparabola}).
The penultimate column, $s_{\mathrm{H.E.S.S.}}$, gives the best-fit energy-scale shift parameter applied to the H.E.S.S.\ data (see Section~\ref{data_analysis_pipeline_2}).
A dash in the $\beta$ column indicates that the spectrum is consistent with a simple power-law form.
All fits assume the EBL model in \cite{2011MNRAS.410.2556D}, used as a baseline.
The final column lists references to H.E.S.S.\ publications reporting observations of the sources during contemporaneous or closely related epochs \cite{2024hegr.confE.171G}.
}
\label{tab:combined_blazar_summary}
\end{table*}


\section{Effect of ALPs decay on the EBL}
\label{alps_decay_theory}

Unlike previous EBL measurements based on blazar spectra \cite{2017A&A...606A..59H}, the formalism introduced here does not aim to reconstruct the total EBL, but instead models a specific ALP-induced contribution to the optical depth on top of established EBL models.
The EBL intensity due to a cosmological population of decaying ALPs is given by \citep{overduin_dark_2004,2022PhRvL.129w1301B,2024PhRvD.110j3501P}
\begin{equation} 
\nu I_{\nu} (\lambda,z) = \frac{\Omega_a\rho_\mathrm{crit,0} c^4}{64\pi} 
\frac{(m_a c^2)^2 g_{a\gamma}^2}{\lambda\, H(z_{\ast})} 
\Theta\!\left(\lambda - \frac{2hc}{m_a c^2}\right),
\label{eq:axiondecayCOSMIC}
\end{equation}
where \( \nu I_\nu(\lambda,z) \) is the specific intensity at observed wavelength \( \lambda \) and redshift \( z \), and \( \lambda \) denotes the observed photon wavelength. 
This expression assumes that ALPs decay at rest, and any effects due to their motion are neglected.
The constants \( c \) and \( h \) are the speed of light and Planck’s constant, respectively, while \( m_a \) and \( g_{a\gamma} \) are the ALP mass and photon coupling. 
Furthermore, \( \Omega_a\rho_{\mathrm{crit},0} \) is the present-day ALP energy density, with \( \Omega_a \) denoting the fractional ALP density and \( \rho_{\mathrm{crit},0} \) the critical density of the Universe today. 
The redshift $z_{\ast}$ corresponds to the epoch at which ALP decay produces photons that are observed with wavelength $\lambda$ at source redshift $z$,
\begin{equation}
z_{\ast} = \frac{m_a c^2}{2}\,\frac{\lambda}{hc}(1+z) - 1,
\label{eq:zstar}
\end{equation}
and therefore depends explicitly on $z$. As a result, the ALP–decay contribution to the EBL is intrinsically redshift-dependent.
The Heaviside function \( \Theta(x) \) ensures that only decays producing photons longer than the rest-frame wavelength \( 2hc/m_a c^2 \) contribute.  

Following \cite{2024PhRvD.110j3501P}, we model the ALP two-photon decay as a $\delta$-function in energy, since the intrinsic fractional width—set by the small non-relativistic velocity dispersion of ALPs—is negligible once integrated over redshift.
The redshift-dependent Hubble parameter is
\begin{equation}
H(z) = H_0 \sqrt{\Omega_m (1+z)^3 + \Omega_\Lambda}, 
\end{equation}
where we adopt \( \Omega_m = 0.315 \), \( \Omega_\Lambda = 0.685 \), and \( H_0 = 67.4~\mathrm{km\,s^{-1}\,Mpc^{-1}} \), consistent with the \textit{Planck}~2018 results~\cite{Planck2018}.
Where individual EBL models are constructed assuming slightly different cosmological parameters, we adopt values consistent with those models. We verified that varying \(H_0\) within the range spanned by these assumptions and by current local and cosmological measurements leads to changes well within the systematic uncertainties and does not affect the resulting constraints.

In this work, we focus on ALP masses in the range $m_a \simeq 2.5$--$20~\mathrm{eV}$, corresponding to decay photons with wavelengths $\lambda_{\rm decay} = 2hc/(m_a c^2) \sim 120 - 1000~\mathrm{nm}$. 
This optical to near-infrared band overlaps with the portion of the EBL that most strongly attenuates TeV gamma rays through $\gamma\gamma$ pair production, for which the cross-section peaks when $\lambda_{\rm EBL} \simeq 1.2 (E_\gamma/\mathrm{TeV}) ~\mu\mathrm{m}$. 
For $m_a < 2.5~\mathrm{eV}$, the decay photons fall predominantly in the infrared regime, contributing mainly at wavelengths that have limited impact on TeV-scale absorption. 
At masses around $m_a \sim 1~\mathrm{eV}$, the resulting effect becomes sufficiently weak that only couplings $g_{a\gamma} \gtrsim 10^{-10}~\mathrm{GeV}^{-1}$ would lead to observable modifications; such couplings are already excluded by constraints from globular cluster stars and other astrophysical probes~\cite{AxionLimits}.

Conversely, for $m_a \gtrsim 20~\mathrm{eV}$, the photons are produced in the far-UV or soft X-ray band ($\lambda \lesssim 60~\mathrm{nm}$), where the Universe is highly opaque due to photoelectric absorption by neutral hydrogen and helium. Such photons do not contribute to the diffuse EBL and are strongly constrained by the measured extragalactic X-ray background and CMB spectral-distortion limits.  Hence, the $2.5$--$20~\mathrm{eV}$ window represents the most physically relevant and observationally accessible mass range for testing the impact of heavy ALP decay on gamma-ray optical depth.

Furthermore, the two-photon decay lifetime of a heavy axion-like particle is
\begin{equation}
\label{eq:Ta_decay}
\tau_{a\gamma\gamma}
    = \frac{64\pi\hbar}{(m_a c^2)^3 g_{a\gamma}^2}
    \simeq 1.3\times10^{24}\ \mathrm{s}
    \left(\frac{m_a c^2}{10~\mathrm{eV}}\right)^{-3}
    \left(\frac{g_{a\gamma}}{10^{-11}\ \mathrm{GeV}^{-1}}\right)^{-2},
\end{equation}
which shows explicitly that the ALP lifetime decreases steeply with increasing mass,
\(\tau_{a\gamma\gamma}\propto m_a^{-3}\) for fixed coupling.  
Thus, heavier ALPs decay more rapidly on cosmological timescales, producing a larger
photon emissivity, and consequently a stronger ALP-induced contribution to the EBL
and to the gamma-ray optical depth.
Using Eq.~\eqref{eq:Ta_decay} with $g_{a\gamma}=10^{-11}\,\mathrm{GeV}^{-1}$, the corresponding lifetimes span from $\tau_{a\gamma\gamma}\sim 1\times10^{25}\,\mathrm{s}$ at $m_a=5~\mathrm{eV}$ to $\tau_{a\gamma\gamma} \sim 2\times10^{23}\,\mathrm{s}$ at $m_a=20~\mathrm{eV}$, still many orders of magnitude longer than the age of the Universe.

\begin{figure*}[tb]
    \centering
    \includegraphics[width=.99\linewidth]{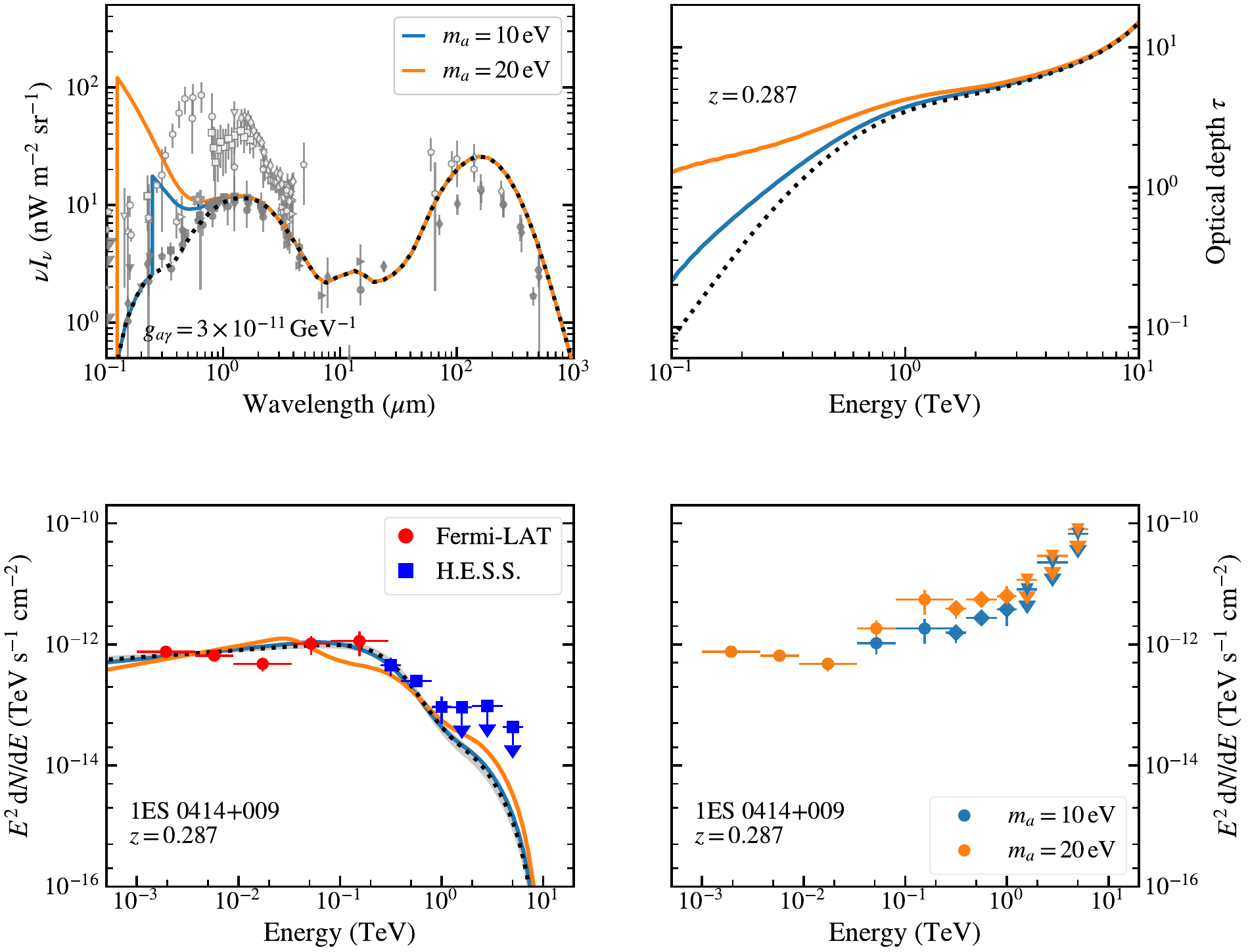}
\caption{
\textbf{Top Left:} The Domínguez EBL model \citep{2011MNRAS.410.2556D} (dotted black), together with two illustrative ALP–decay contributions for axion masses of $m_a=10$\,eV (blue) and $m_a=20$\,eV (orange), assuming a fixed coupling of $g_{a\gamma} = 3\times10^{-11}\,\mathrm{GeV}^{-1}$ (solid lines).
Observed EBL lower and upper limits are shown as filled and open symbols, respectively \citep{Biteau_The_MM_EGAL_spectrum,2018ApSpe..72..663H}.
\textbf{Top Right:} Gamma-ray optical depths as a function of energy at $z = 0.287$, corresponding to the redshift of 1ES~0414+009, derived from the same EBL models shown in the top left panel using \texttt{ebltable}.
The blue and orange curves correspond to ALP masses of $m_a=10$\,eV and $m_a=20$\,eV, respectively, while the dotted black curve shows the no-ALP Domínguez EBL model.
\textbf{Bottom Left:} Best-fit gamma-ray spectra of 1ES~0414+009 obtained using the EBL models shown in the top panels (the Domínguez EBL alone and with illustrative ALP--decay contributions).
The flux points show the observed \textit{Fermi}--LAT and H.E.S.S.\ spectrum and are computed under the no-ALP hypothesis; recomputing the flux points using the ALP-modified optical depths shown here produces negligible changes. 
\textbf{Bottom Right:} The gamma-ray flux points corrected for EBL absorption while retaining the ALP-induced modification to the optical depth, corresponding to the same ALP scenarios shown in the other panels.
Circles denote \textit{Fermi}--LAT points and diamonds/triangles denote H.E.S.S.\ points (detections and upper limits, respectively), with colors indicating the two ALP masses.}
\label{fig:alpDecay_example}
\end{figure*}

To illustrate the impact of ALP decay on the EBL and subsequent gamma-ray attenuation, Figure~\ref{fig:alpDecay_example} presents an example based on the blazar 1ES~0414+009. 
In the top left panel, the dotted curve shows the baseline EBL model of \cite{2011MNRAS.410.2556D}, based solely on known astrophysical sources and consistent with the lower observational limits (filled symbols).
Two illustrative ALP decay components are shown as solid lines, computed from Eq.~\eqref{eq:axiondecayCOSMIC} for  masses of $m_a=10$\,eV and $m_a=20$\,eV, assuming a fixed coupling of $g_{a\gamma} = 3\times10^{-11}\,\mathrm{GeV}^{-1}$ and \( \Omega_a=\Omega_{\rm DM}\).
These cases demonstrate how ALP decays can enhance the EBL intensity and thereby increase gamma-ray optical depth.
The top right panel shows the corresponding optical depths, calculated using \texttt{ebltable} \citep{manuel_meyer_2022_7312062}. 
The optical depth $\tau(E,z)$ is computed following \cite{2015ApJ...812...60B}:
\begin{equation}
    \tau(E,z) = \int_0^z \frac{dl}{dz'} dz' \int_{-1}^{+1} d\mu \, \frac{1-\mu}{2} \int_{\epsilon_\mathrm{thr}}^{\infty} d\epsilon \, n(\epsilon,z') \, \sigma_{\gamma\gamma}\!\left(s=2E\epsilon(1-\mu)\right),
\end{equation}
where $n(\epsilon,z')$ is the proper number density of EBL photons of energy $\epsilon$ at redshift $z'$, $\sigma_{\gamma\gamma}$ is the pair-production cross section, and $dl/dz'$ is the cosmological line element.
In our analysis, the total optical depth is taken as the sum of the conventional EBL contribution and the additional effective component induced by ALP-decay photons, ensuring both conventional $\gamma\gamma$ absorption and the decay-induced component are consistently included.
The ALP–decay contribution to the EBL and the resulting optical depth are redshift-dependent; the curves shown in Figure~\ref{fig:alpDecay_example} correspond specifically to the redshift of 1ES~0414+009.

For ALPs with $m_a \sim 10~\mathrm{eV}$, the added optical depth peaks in the $\sim100~\mathrm{GeV}$--$1~\mathrm{TeV}$ range. 
Even in extreme cases that violate EBL constraints, the increase in optical depth typically remains within an order of magnitude.
The bottom left panel of Figure~\ref{fig:alpDecay_example} shows the resulting best-fit spectra under the three optical depth models shown in the top panels, illustrating how ALP-induced absorption modifies the inferred spectral shape.
The flux points in the bottom left panel of Figure~\ref{fig:alpDecay_example} for both \textit{Fermi}--LAT and H.E.S.S.\ data are computed under the no-ALP assumption; using ALP-modified optical depth models produces negligible differences in the flux point levels.
To further illustrate this point, the bottom right panel of Figure~\ref{fig:alpDecay_example} shows the corresponding flux points after correcting for EBL absorption while retaining the ALP-induced modification to the optical depth. This highlights that, despite the significant differences in optical depth between the models, the reconstructed intrinsic fluxes remain relatively similar. This behaviour reflects the fact that changes in absorption can be largely compensated by adjustments in the fitted intrinsic spectrum. As a result, even sizable variations in optical depth lead to only modest changes in the inferred emission.
We emphasise that the flux points are shown for visualisation only, while the spectral fits are performed directly in counts space using the full IRFs.

Figures~\ref{fig:no_alps_spectra_a} and~\ref{fig:no_alps_spectra_b} show the baseline fits for all sources and time segments, obtained using the EBL model of \cite{2011MNRAS.410.2556D}. 
For each interval, we perform a joint spectral fit across \textit{Fermi}--LAT and H.E.S.S.\ data using the binned Poisson likelihood implemented in \texttt{Gammapy}, combining both instruments into a unified dataset and maximising the joint likelihood directly, using the appropriate IRFs that account for energy resolution and energy migration (see later for details).
This approach ensures that full spectral information is used without relying solely on precomputed flux points.

For sources with multiple variability-based segments (e.g.\ H2356-309 and PKS~2155$-$304), we show the best-fit models for each interval separately. 
Both simple power-law and log-parabola spectral forms are tested as models for the intrinsic source spectra:
\begin{equation}
  \frac{dN}{dE} = N_{0} \left(\frac{E}{E_{0}}\right)^{-\gamma},
  \label{eq:powerlaw}
\end{equation}
and
\begin{equation}
  \frac{dN}{dE} = N_{0} \left(\frac{E}{E_{0}}\right)^{-\alpha - \beta \ln \left(\frac{E}{E_{0}}\right)},
  \label{eq:logparabola}
\end{equation}
where $N_{0}$ is the flux normalisation (in photons\,cm$^{-2}$\,s$^{-1}$\,TeV$^{-1}$), 
$E_{0}$ is the fixed pivot energy, 
$\gamma$ and $\alpha$ are the spectral indices, 
and $\beta$ quantifies the curvature in the log-parabola model. 
We note that an exponential cutoff represents an alternative physically motivated parameterisation at the highest energies. We tested such models and found that they are not statistically preferred over the power-law or log-parabola descriptions for the spectra analysed here. Because both a cutoff and additional absorption suppress the spectrum at high energy, they are partially degenerate; we therefore do not include exponential cutoffs in the full mass--coupling scan.

Our spectral analysis is performed using \texttt{Gammapy}, following the procedures established for joint multi-instrument likelihood studies \citep[e.g.][]{Aharonian:2023hessIGMF,2024A&A...686A.308A}. For each variability-based time segment, we construct a unified \texttt{Gammapy} \texttt{Datasets} object that combines the \textit{Fermi}--LAT and H.E.S.S.\ data. The IRFs, in particular the point-spread function and energy-dispersion kernels, are exported from the event-level information and attached separately for each segment, ensuring that temporal variations in the instrument performance are treated consistently. All background and diffuse parameters are fixed to the values obtained in the segment-level analyses, so that only the spectral parameters of the central blazar remain free. The joint spectral reconstruction is obtained by maximising the binned Poisson likelihood across both instruments simultaneously.

In addition to statistical uncertainties, we explicitly account for a possible residual H.E.S.S.\ energy-scale bias. Stochastic effects are modelled through the energy-dispersion kernels in the IRFs. To capture a coherent shift of the absolute energy scale, we introduce a nuisance parameter $s$ that rescales the reconstructed energies as $E \rightarrow E(1+s)$, applied only to the H.E.S.S.\ datasets. This parameter is constrained with a Gaussian prior with a width corresponding to the $\sim10$--15\% H.E.S.S.\ energy-scale uncertainty, and is profiled over together with the spectral parameters in the likelihood fit. Such a smooth energy rescaling is largely degenerate with intrinsic spectral shape and does not impact the redshift-dependent absorption expected from an ALP-induced optical-depth contribution, rendering our limits conservative. Systematic uncertainties in the \textit{Fermi}--LAT energy range are negligible for this analysis.

For each variability segment, we perform both spectral fits within the joint
\texttt{Gammapy} likelihood framework and select the log-parabola only when it provides a
statistically significant improvement over the power-law, as quantified by the
likelihood-ratio test for one additional degree of freedom, corresponding to a $3\sigma$ significance threshold. Otherwise, the simple
power-law model is retained. For joint \textit{Fermi}--LAT and H.E.S.S.\ fits, the reference energy $E_{0}$ is fixed to the value reported in the \textit{Fermi}--LAT 4FGL-DR4 catalogue for each source, whereas for H.E.S.S.-only fits it is determined by the H.E.S.S.\ data and typically lies in the VHE range. The corresponding best-fit spectral parameters for all sources and time intervals are summarised in Table~\ref{tab:combined_blazar_summary}.

\renewcommand{\thefigure}{\arabic{figure}a}
\begin{figure*} 
    \centering
    \includegraphics[width=0.32\linewidth]{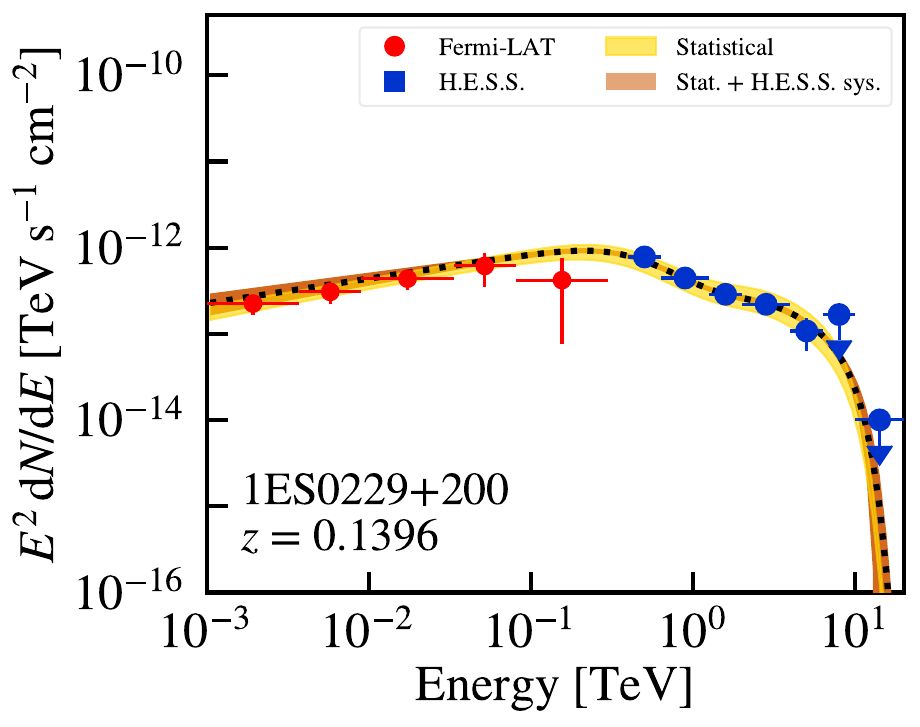}
\includegraphics[width=0.32\linewidth]{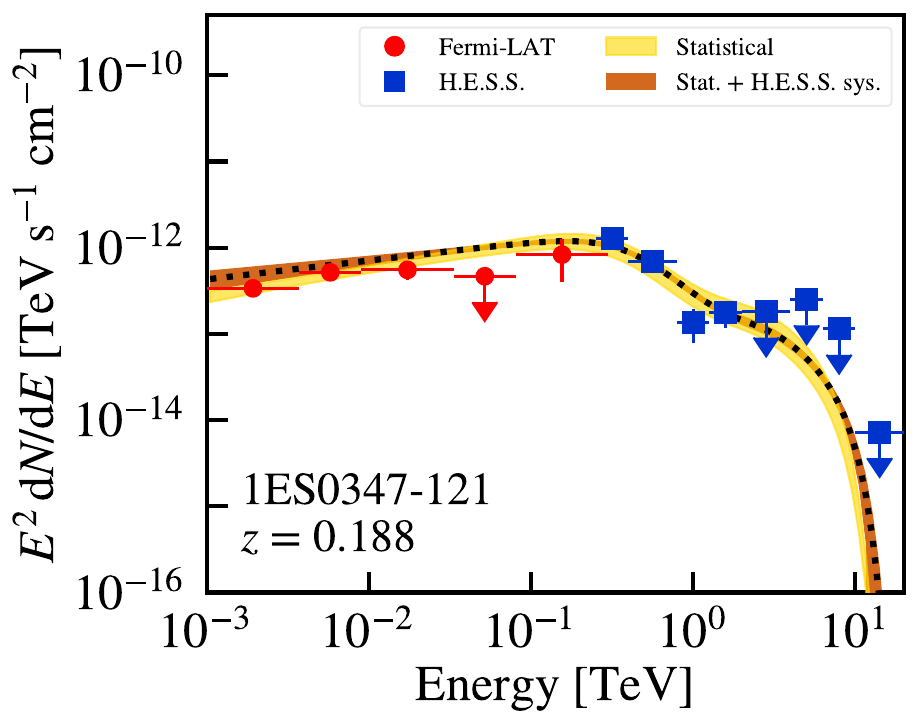}
    \includegraphics[width=0.32\linewidth]{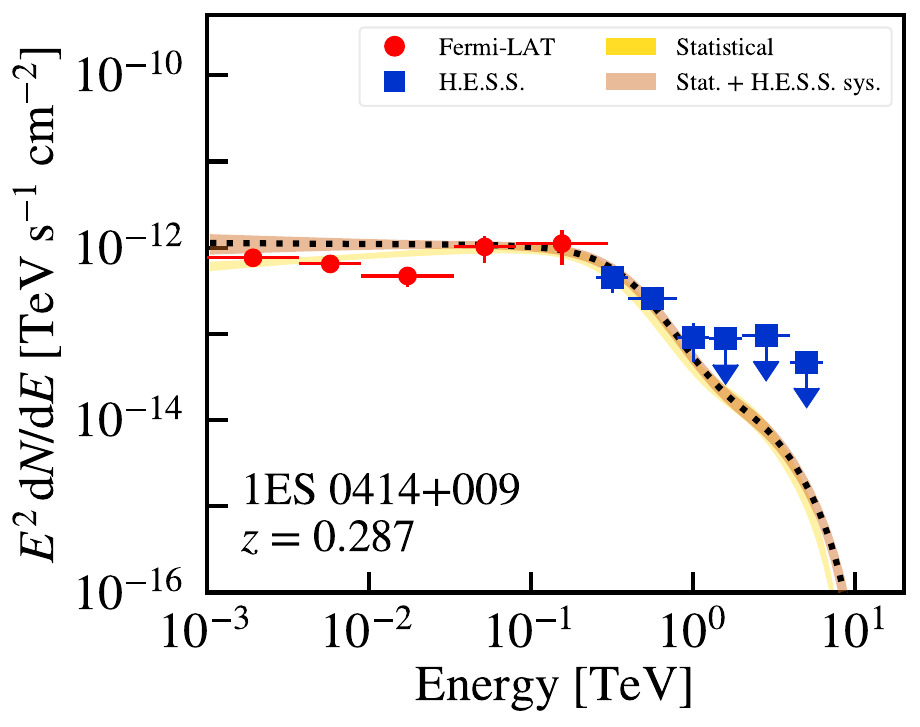}
    \includegraphics[width=0.32\linewidth]{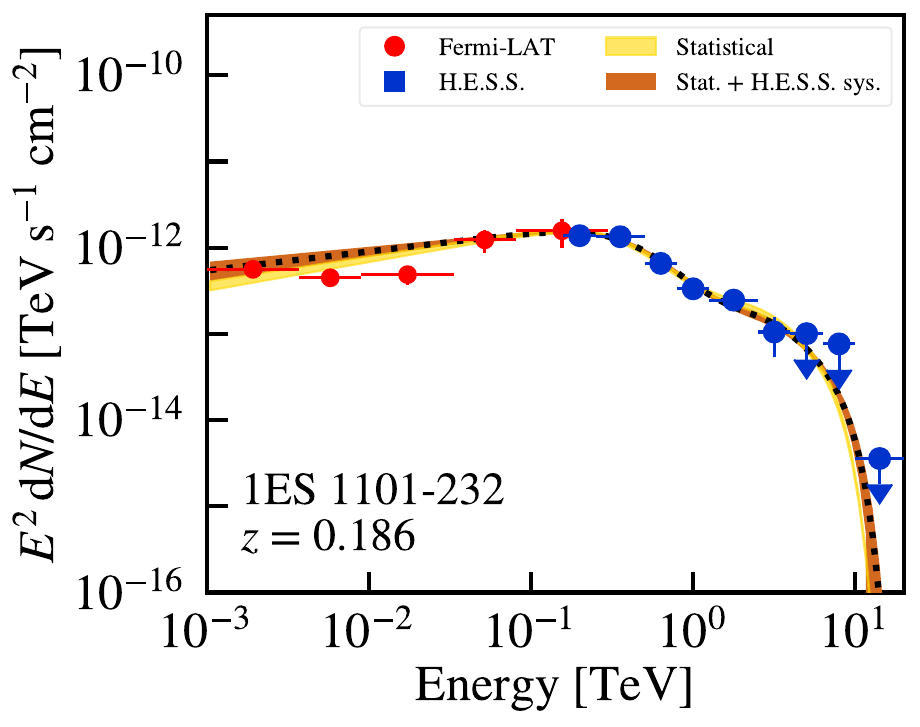}
    \includegraphics[width=0.32\linewidth]{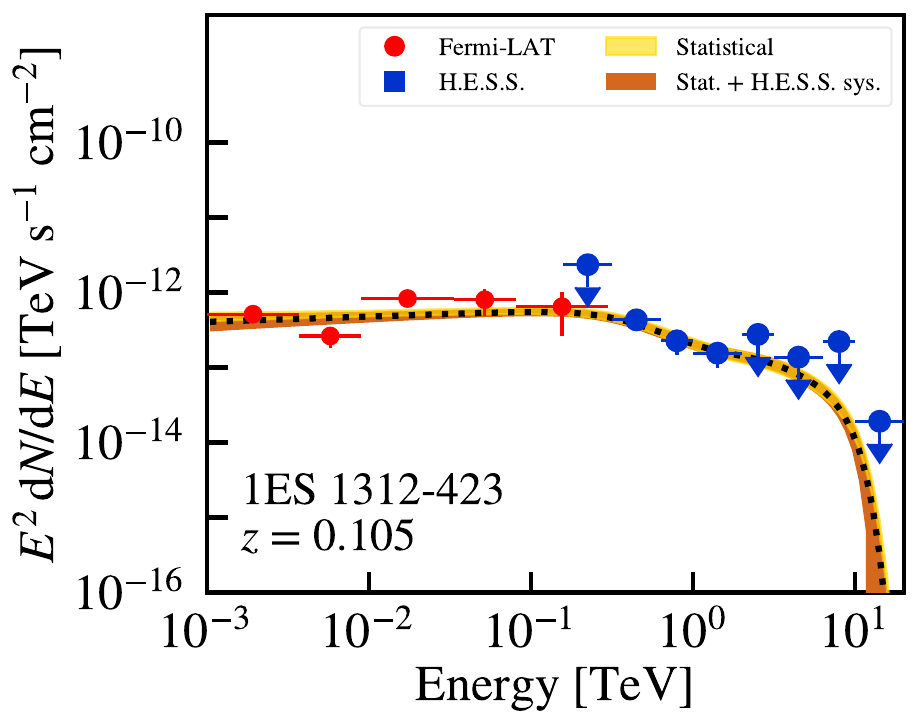}
     \includegraphics[width=0.32\linewidth]{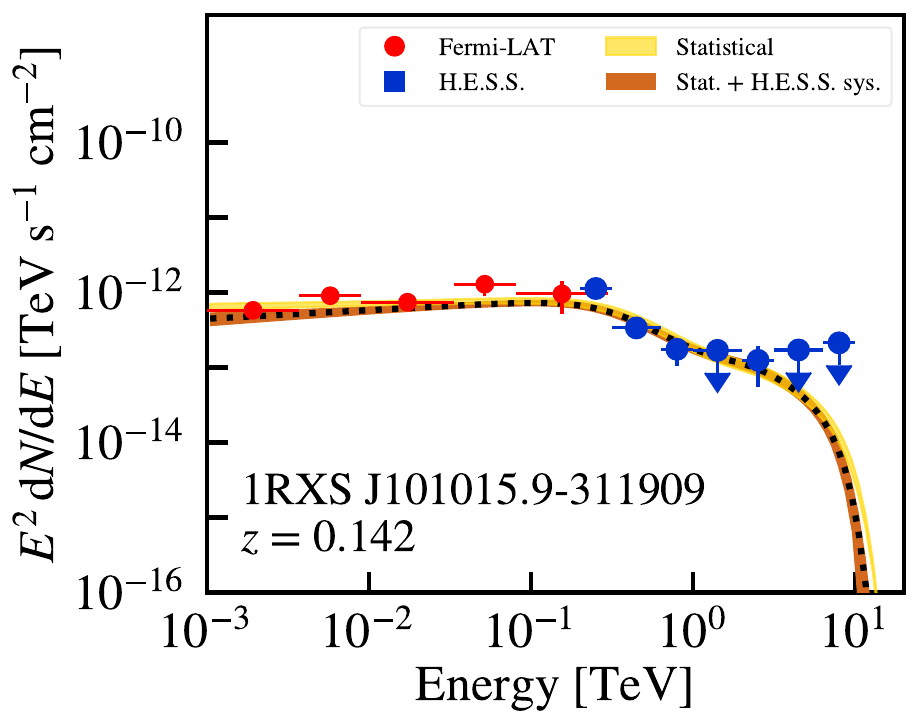} 
    \includegraphics[width=0.32\linewidth]{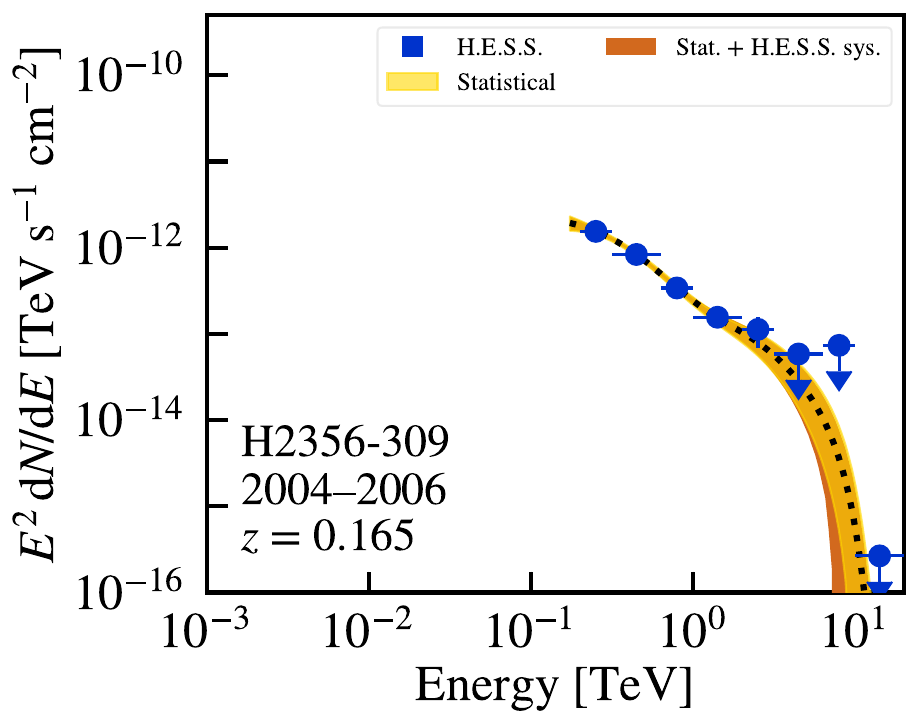}
     \includegraphics[width=0.32\linewidth]{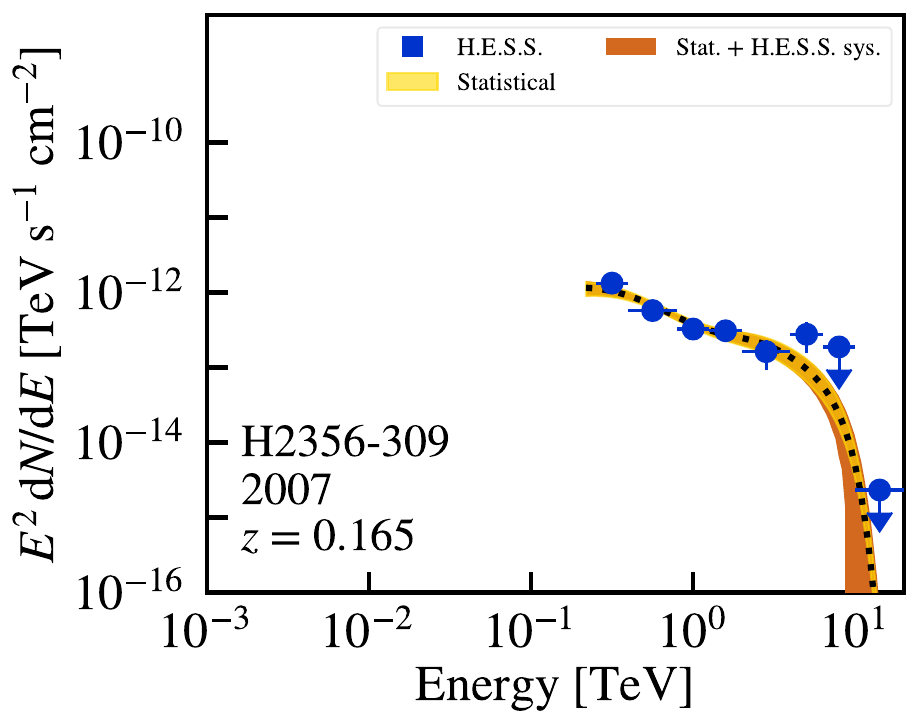}
     \includegraphics[width=0.32\linewidth]{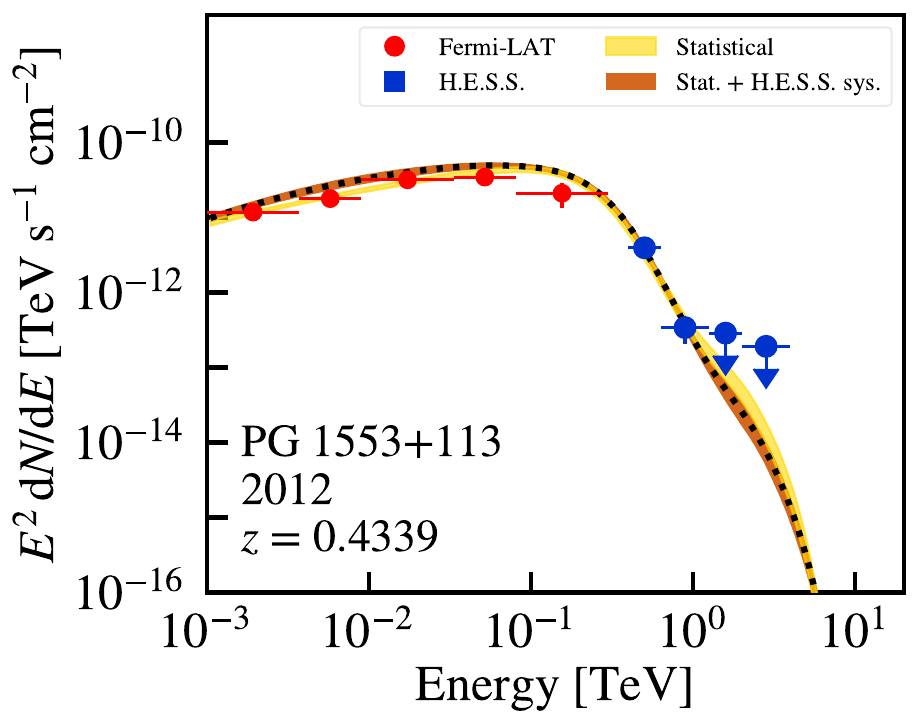}

   \caption{Best-fit spectra for the first set of sources and  time intervals investigated, obtained using the no-ALP scenario, as described in Section~\ref{alps_decay_theory}. The corresponding best-fit parameters are listed in Table~\ref{tab:combined_blazar_summary}.}
     \label{fig:no_alps_spectra_a}
\end{figure*}

\addtocounter{figure}{-1}
\renewcommand{\thefigure}{\arabic{figure}b}
\begin{figure*} 
    \centering
    \includegraphics[width=0.32\linewidth]{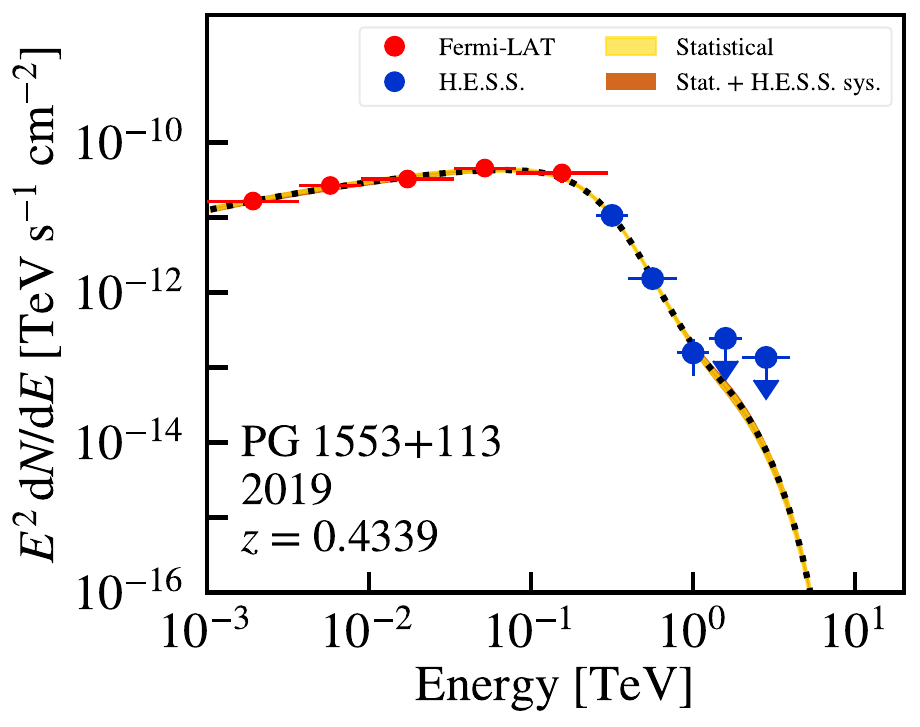}
    \includegraphics[width=0.32\linewidth]{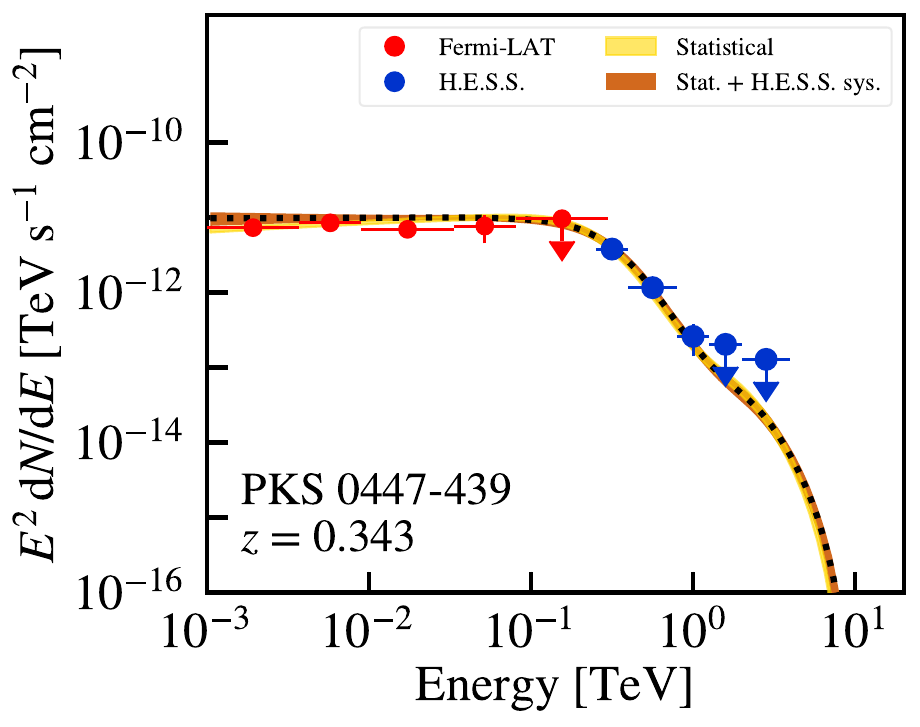}
    \includegraphics[width=0.32\linewidth]{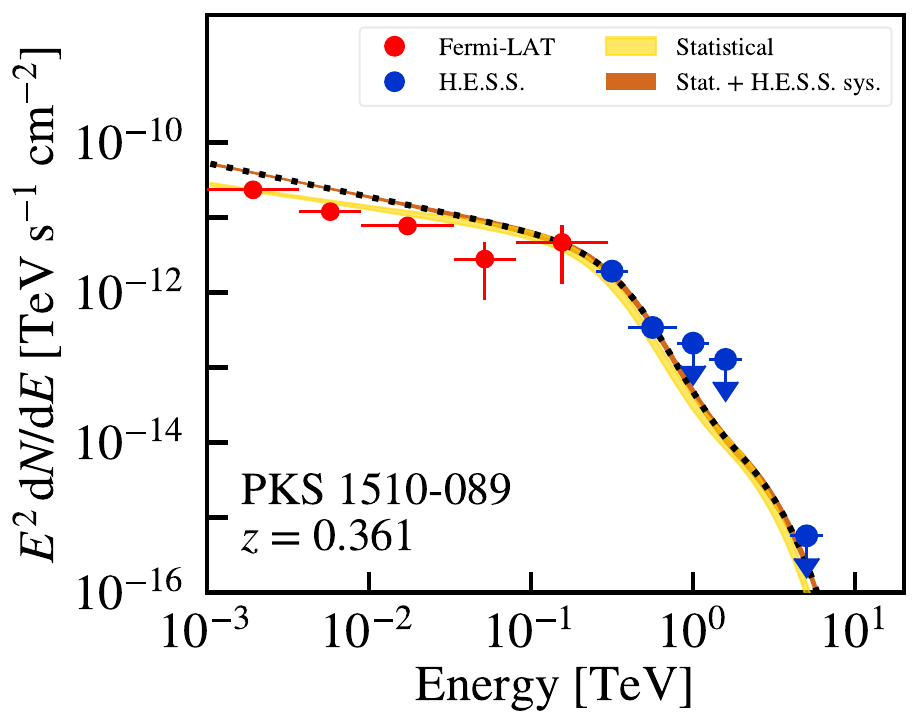}
    \includegraphics[width=0.32\linewidth]{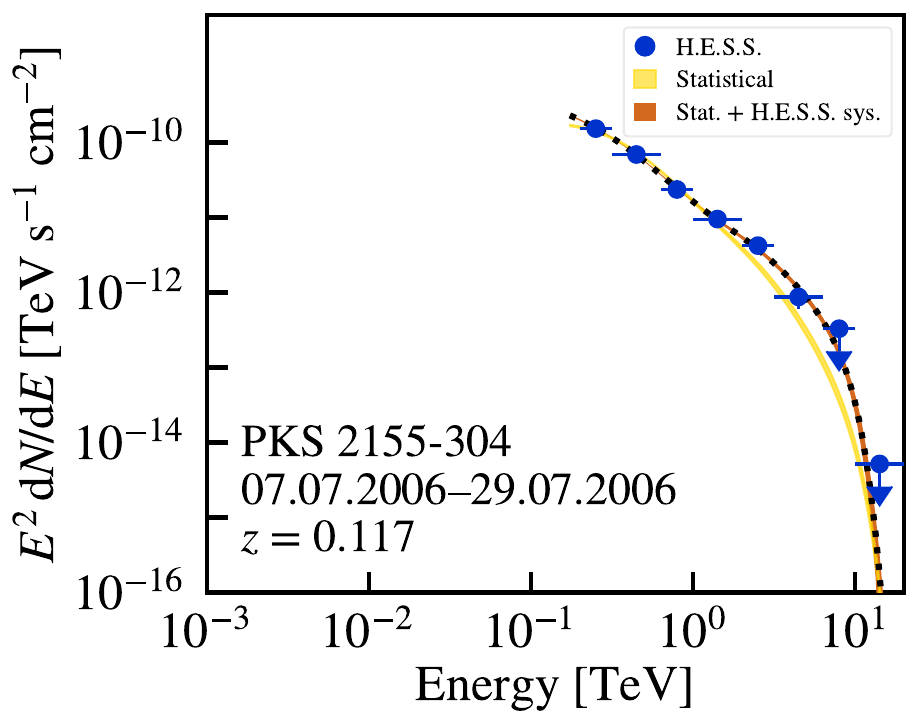}
    \includegraphics[width=0.32\linewidth]{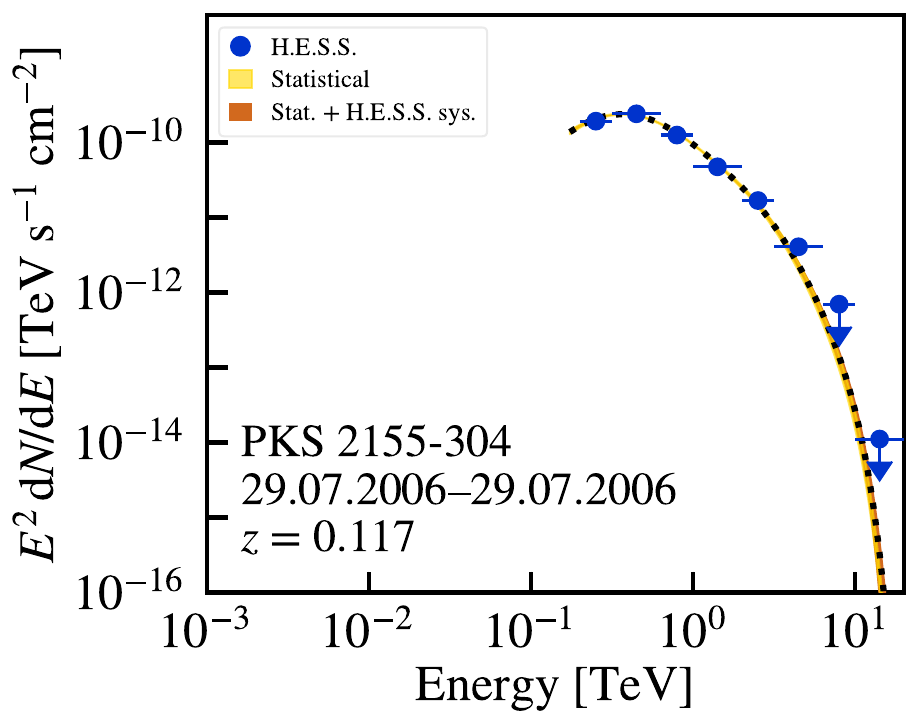}
    \includegraphics[width=0.32\linewidth]{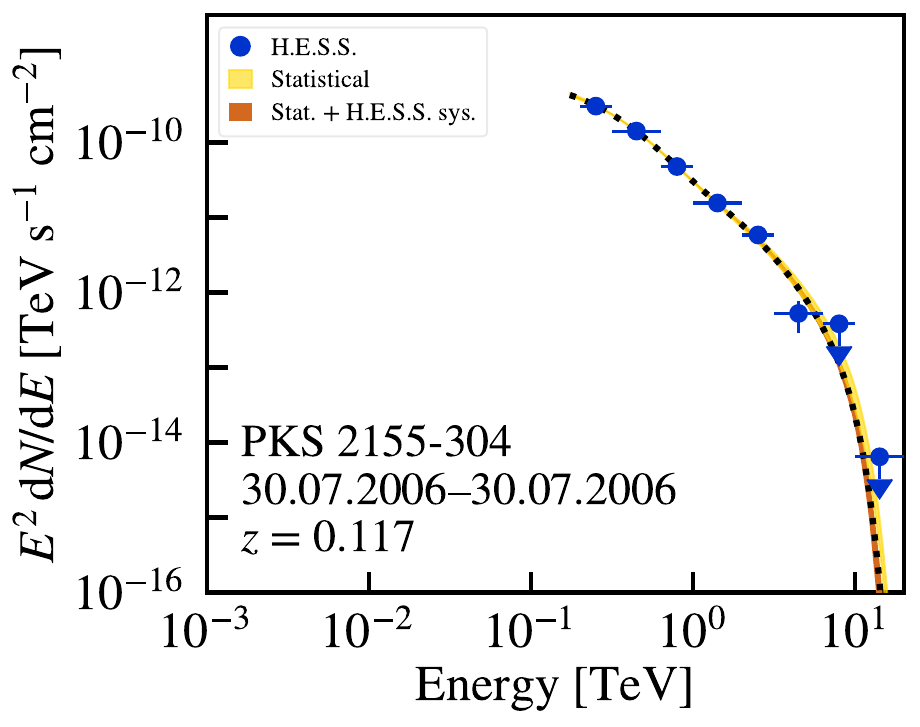}
    \includegraphics[width=0.32\linewidth]{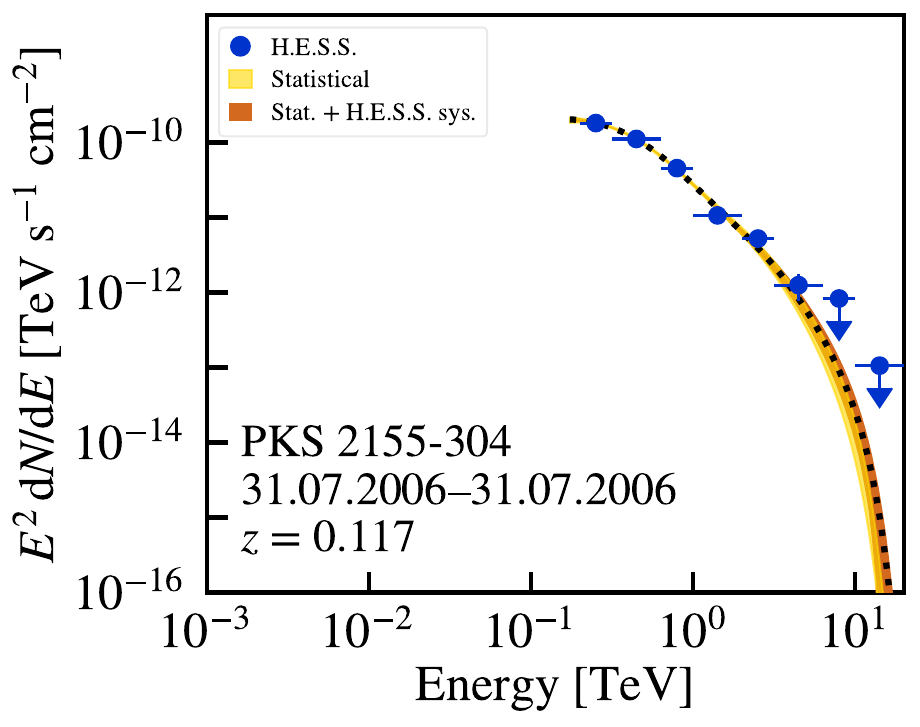}
    \includegraphics[width=0.32\linewidth]{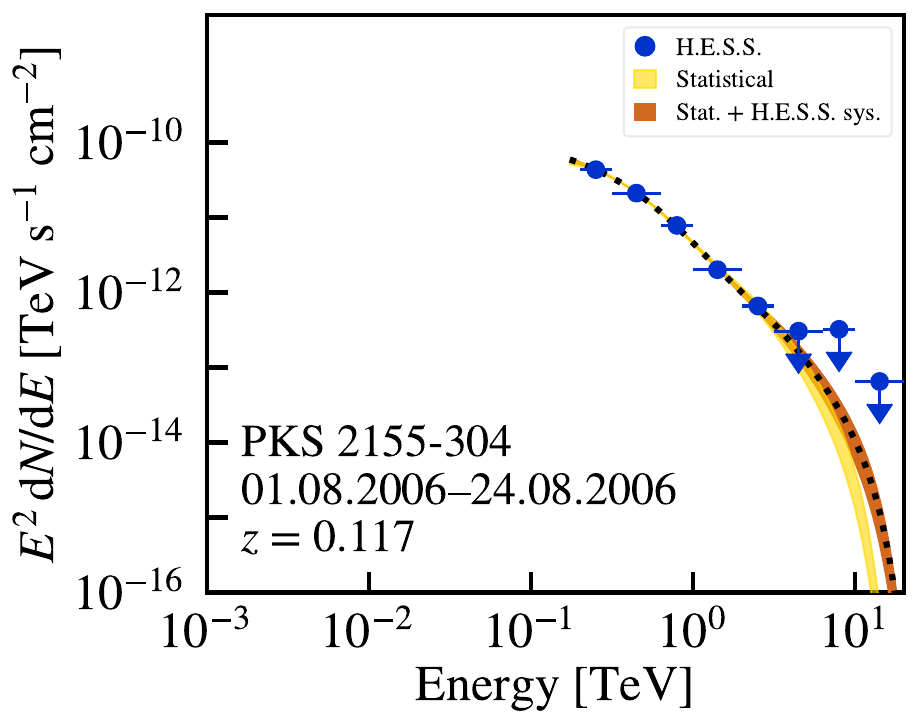}
   \includegraphics[width=0.32\linewidth]{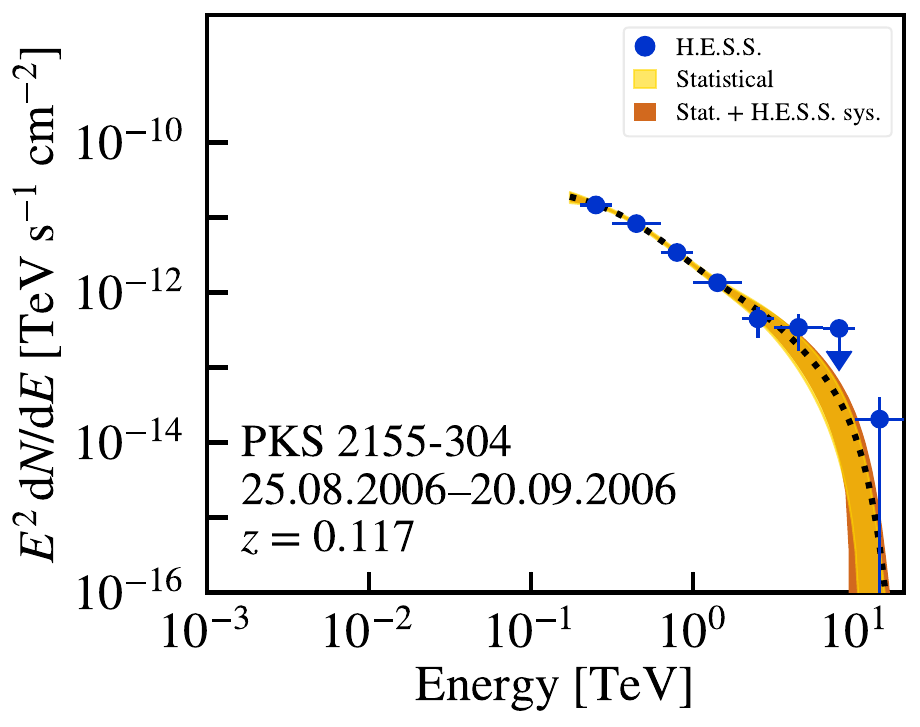}
    \caption{Same as Fig.~\ref{fig:no_alps_spectra_a}.}
    \label{fig:no_alps_spectra_b}
\end{figure*}

\renewcommand{\thefigure}{\arabic{figure}}

\vspace{0.3cm}
\section{Results and Discussion}

A joint spectral analysis of the combined \textit{Fermi}--LAT and H.E.S.S.\ data is performed for each source and time interval to test for potential ALP-induced modifications to the EBL attenuation.  
The null (no-ALP) fits serve as the reference model for each segment (Figures~\ref{fig:no_alps_spectra_a} and \ref{fig:no_alps_spectra_b}).  
We then introduce the ALP-decay contribution described in Eq.~\eqref{eq:axiondecayCOSMIC} and repeat the spectral fits over the ALP mass range motivated in Section~\ref{alps_decay_theory}, scanning photon--ALP coupling constants \(g_{a\gamma}\).
For fixed values of \(m_a\) and \(g_{a\gamma}\) and each spectrum $i$ with data $D_i$ this corresponds to maximising the joint \textit{Fermi}-LAT and H.E.S.S. log-likelihood in terms of nuisance parameters $\boldsymbol{\theta}_i$ (spectral and background parameters),
\begin{equation}
    \ln\mathcal{L}(m_a, g_{a\gamma}, \boldsymbol{\theta}_i, s_i \mid D_i)
    =
    \sum_{j}\ln\mathcal{L}(m_a, g_{a\gamma}, \boldsymbol{\theta}_i \mid D_{\mathit{Fermi},ij})
    +
    \sum_{j}\ln\mathcal{L}(m_a, g_{a\gamma}, \boldsymbol{\theta}_i, s_i \mid D_{\mathrm{H.E.S.S.},ij})
    -
    \frac{s_i^2}{2\sigma_s^2},
\end{equation}
where the sum over $j$ runs over the energy bins for both \textit{Fermi}~LAT and H.E.S.S., $s_i$ denotes the nuisance parameter describing a coherent H.E.S.S.\ energy-scale shift, constrained by a Gaussian prior of width $\sigma_s$, and both likelihoods follow Poissonian statistics. The likelihood for the H.E.S.S. observations takes into account both the On region containing the source as well as Off regions to estimate the background, and is given by 
\begin{equation}
  \ln\mathcal{L}(m_a, g_{a\gamma}, \boldsymbol{\theta}_i, s_i \mid D_{\mathrm{H.E.S.S.},ij})
  =
  N_{\mathrm{On}, ij} \ln (\mu_{ij}(s_i) + b_{ij})
  +
  N_{\mathrm{Off},ij} \ln (b_{ij} / \alpha)
  -
  \mu_{ij}(s_i)
  -
  \left(1 + \frac{1}{\alpha}\right)b_{ij},
\end{equation}

where $N_\mathrm{On}$ and $N_\mathrm{Off}$ are the counts in the On and Off regions, respectively, $\mu_{ij}(s_i)$ are the expected number of counts from the source computed using the IRFs (effective area and energy-dispersion/migration matrices) and after rescaling the reconstructed H.E.S.S.\ energies as $E \rightarrow E(1+s_i)$, depending on the spectral and ALP parameters as well as the EBL density, and $b_{ij}$ are the expected counts from the background.

For a fixed ALP mass, we then construct the profile log-likelihood ratio as a sum over all spectra,
\begin{equation}
\lambda(m_a, g_{a\gamma}) = -2\sum_i \left[
\ln\mathcal{L}(m_a, g_{a\gamma}, \hat{\boldsymbol{\theta}}_i \mid D_i)
-
\ln\mathcal{L}(m_a, \hat{g}_{a\gamma}, \hat{\boldsymbol{\theta}}_i \mid D_i)
\right],
\label{eq:lambda}
\end{equation}
where $\hat{\vphantom{\rule{1pt}{9pt}}\smash{\hat{\boldsymbol{\theta}}}}$
 denotes the spectral and background parameters maximising the likelihood for a
fixed ALP coupling strength, whereas $\hat{\boldsymbol{\theta}}$
 and $\hat{g}_{a\gamma}$ are the parameters that
maximise $\mathcal{L}$ unconditionally.

\begin{figure*} 
    \includegraphics[width=\linewidth]{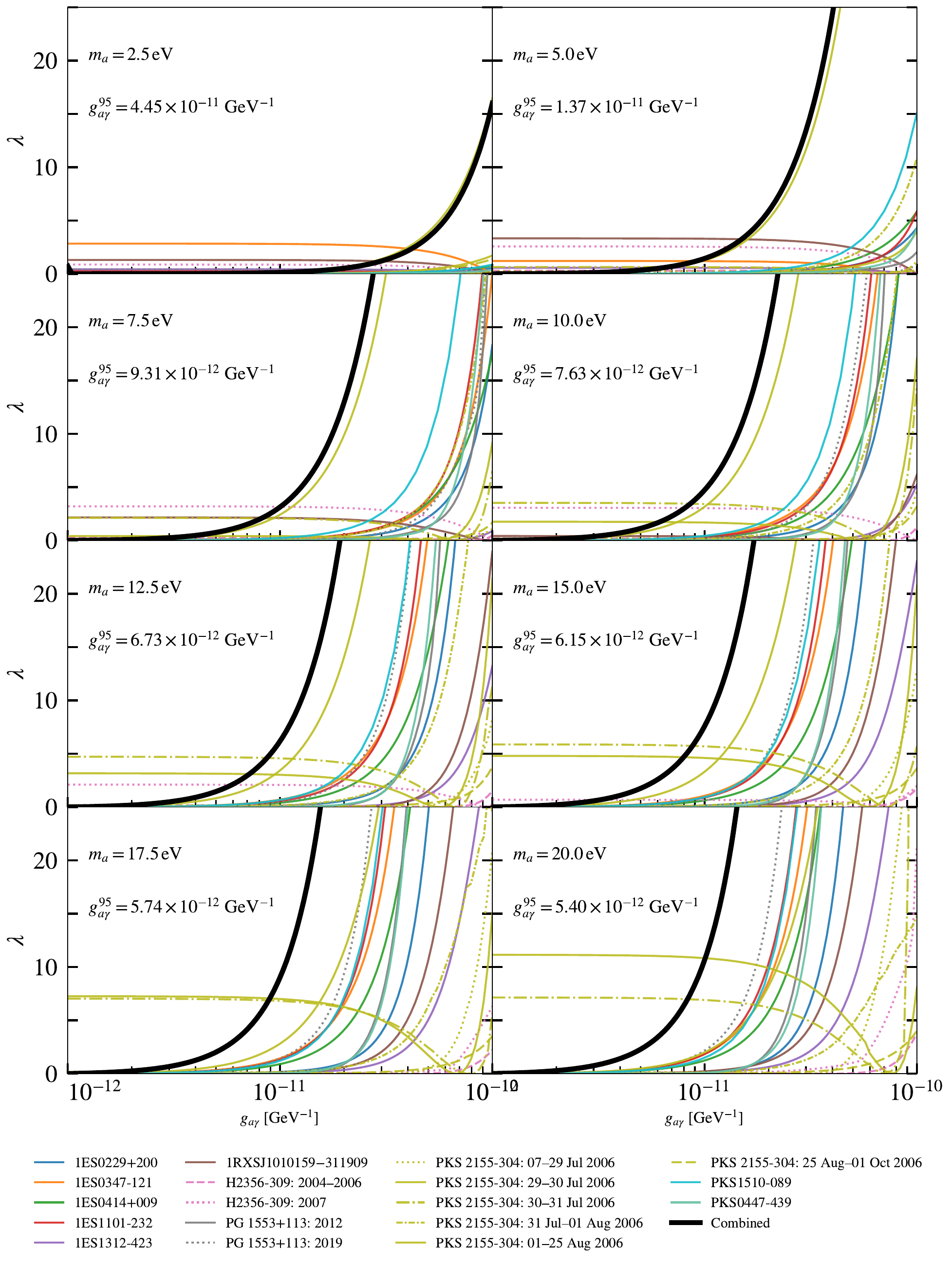}
   \caption{The profile likelihood, \(\lambda(m_a, g_{a\gamma})\), as a function of the photon--ALP coupling constant \(g_{a\gamma}\) for different fixed ALP masses \(m_a\).  
Each panel corresponds to a specific mass, as labelled.  
Here \(\lambda\) is defined in Eq.~\eqref{eq:lambda} as the difference in \(C\)-statistic relative to the best fit at fixed \(m_a\), and the curves are obtained from the joint fit to the combined \textit{Fermi}--LAT and H.E.S.S.\ data, with contributions summed over all time segments.  
The black curve shows the total \(\lambda\).  
The value of \(g_{a\gamma}^{95}\), corresponding to the point where \(\lambda=2.71\) (95\%~C.L., one-sided; one parameter of interest), is indicated in each panel.}

    \label{fig:TS_gag_ma:all}
\end{figure*}

\begin{figure*}
    \centering
    \includegraphics[width=\linewidth]{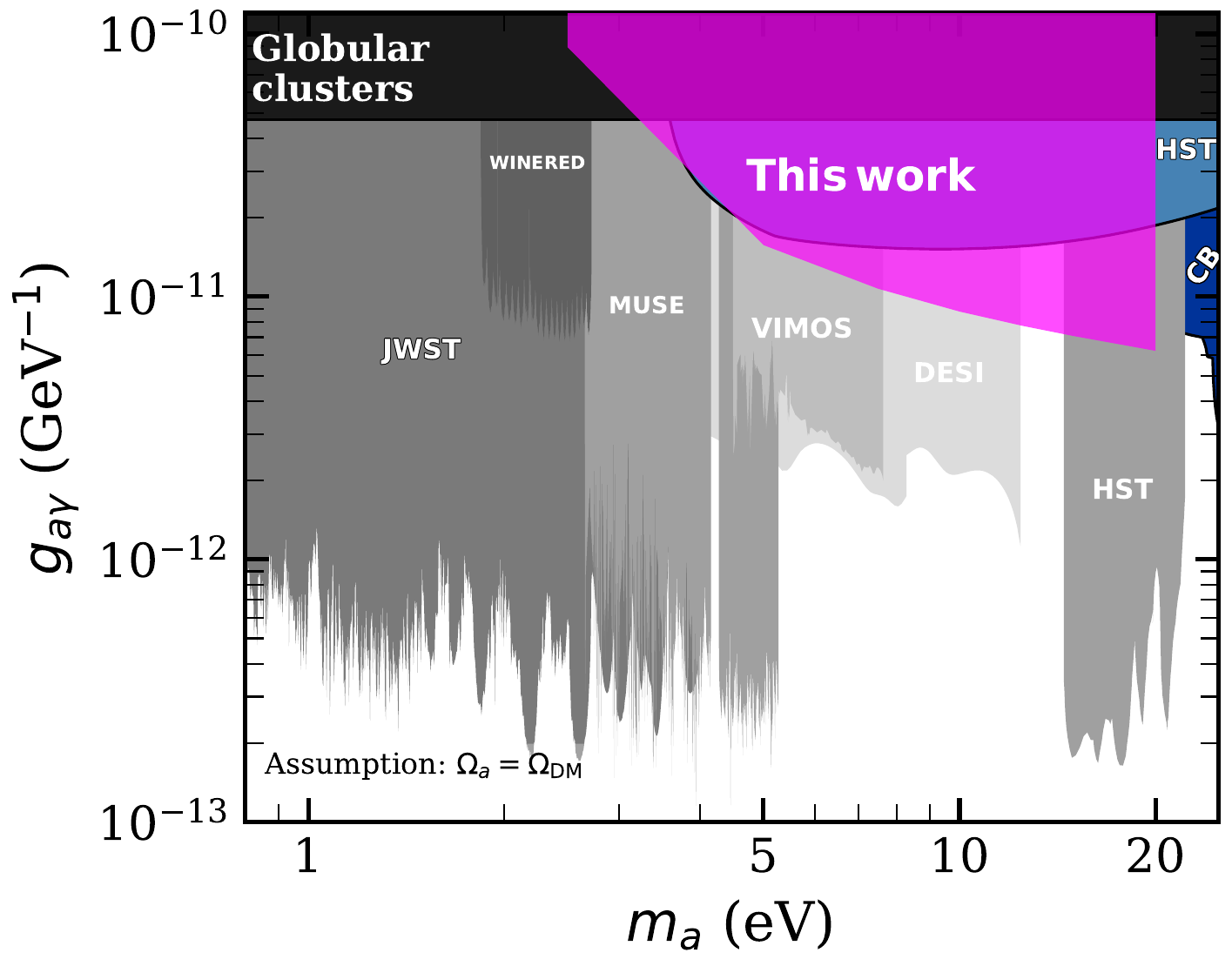}
    \caption{Constraints on the ALP parameter space in the \((m_a, g_{a\gamma})\) plane, under the assumption $\Omega_a=\Omega_{\rm DM}$.  
    The shaded regions show existing limits from astrophysical and laboratory experiments, compiled in the \texttt{AxionLimits} repository~\cite{AxionLimits}.  
    The purple contour indicates the parameter space excluded at 95\%~C.L.\ by this work, based on combined \textit{Fermi}--LAT and H.E.S.S.\ gamma-ray observations.  
    Our results provide new constraints in the low-coupling regime, with the most stringent limits around \(m_a \sim 20\,\mathrm{eV}\), where the impact of ALP decay on gamma-ray optical depth is strongest.}
    \label{fig:limits}
\end{figure*}

By construction, \(\lambda=0\) at the best-fit value of \(g_{a\gamma}\), and larger values of $\lambda$ indicate increasingly disfavoured couplings.  
Since our goal is to set exclusion limits, we follow the dark matter search convention and use the profile likelihood \(\lambda\) [Eq.~\eqref{eq:lambda}] to construct confidence intervals.  
We adopt the criterion \(\lambda = 2.71\), corresponding to a 95\% confidence level (C.L.; one-sided, one parameter of interest).  
For each ALP mass, we extract the coupling where \(\lambda=2.71\); we denote this by \(g_{a\gamma}^{95}(m_a)\equiv\{\,g_{a\gamma}:\lambda(m_a,g_{a\gamma})=2.71\,\}\), and use these points to construct the 95\%~C.L.\ exclusion contour, under the assumption $\Omega_a=\Omega_{\rm DM}$.  
Figure~\ref{fig:TS_gag_ma:all} displays the resulting likelihood profiles $\lambda(m_a, g_{a\gamma})$ for several representative values of $m_a$.
We note that for the individual curves the minima of the C-statistic (i.e. the maximum of the log-likelihood) are subtracted for each dataset, whereas for the combined curve the overall maximum is subtracted; this can lead to apparent offsets between the individual and combined profiles.
The resulting one-sided 95\%~C.L.\ upper limits on the photon--ALP coupling $g_{a\gamma}$ are shown in Figure~\ref{fig:limits}.
The strongest constraints are obtained at the highest ALP masses probed in our scan, where the much shorter decay lifetime (see Eq.~\ref{eq:Ta_decay}) leads to a larger ALP-induced photon emissivity and hence a stronger impact on the gamma-ray optical depth. In our analysis, the strongest constraints are obtained around $m_a \sim 15$--$20~\mathrm{eV}$, reaching $g_{a\gamma}^{95} \simeq 7 \times 10^{-12}\ \mathrm{GeV}^{-1}$.

To identify which observations drive the combined likelihood, we examined the contribution of individual time segments to $\lambda$.  
Across the mass scan, the combined 95\%~C.L.\ upper limit \(g_{a\gamma}^{95}\) is dominated by the PKS~2155\textendash304 flaring segments—particularly the intervals 01\textendash25~Aug~2006 and 30\textendash31~Jul~2006—which contribute \(\sim 7\text{--}18\) and \(\sim 6\text{--}7\), respectively, to the summed profile likelihood (i.e.\ $\lambda$) for \(m_a=7.5\text{--}20\,\mathrm{eV}\).  
At \(m_a=5\,\mathrm{eV}\), the leading contribution comes from H2356\textendash309 (2007), with 1RXS~J101015.9\textendash311909 and PKS~1510\textendash089 subleading, while PG~1553+113 (2012) becomes relevant only at higher masses (e.g., \(\lesssim 1.3\) at \(m_a=20\,\mathrm{eV}\)).  
These dominant contributions vary with ALP mass and dataset and are not aligned across sources.
Furthermore, the smaller spread between models at $m_a = 5$ eV arises because the corresponding decay photons fall in the optical/near-UV wavelength range, where current EBL models are tightly constrained by observational data, leading to similar predictions. In contrast, for $m_a \sim 20$ eV the decay photons probe the far-UV regime, where the EBL is less well constrained, resulting in larger differences between models and hence in the inferred optical depth and limits.

It is worth noting that, while a small number of individual flares or time segments show a mild preference for a nonzero ALP contribution (e.g., the August~2006 flare of PKS~2155\textendash304), the vast majority of segments—and the overall combined analysis—are fully consistent with the null hypothesis.
To scrutinise these outlier segments, Figure~\ref{fig:pks2155_seds} in Appendix~\ref{Appendix} presents SEDs for the two PKS~2155\textendash304 episodes with the largest contributions, overlaid with the null (no-ALP) and best-fit ALP models for \(m_a=7.5\text{--}20\,\mathrm{eV}\).  
Notably, the PKS~2155\textendash304 flares and the H2356\textendash309 episode were observed before the launch of \textit{Fermi}--LAT and therefore rely exclusively on H.E.S.S.\ data.  
This lack of contemporaneous \textit{Fermi}--LAT coverage means that the intrinsic spectrum is not independently constrained at lower energies. As a result, curvature that would ordinarily be fixed by GeV observations can be absorbed into the optical-depth term, artificially enhancing the apparent preference for an ALP component. This behaviour also manifests in Figure~\ref{fig:TS_gag_ma:all}, where some individual likelihood curves show decreasing trends with $g_{a\gamma}$ and contribute only weakly to the combined profile likelihood.

\begin{figure}
    \centering
    \includegraphics[width=0.72\textwidth]{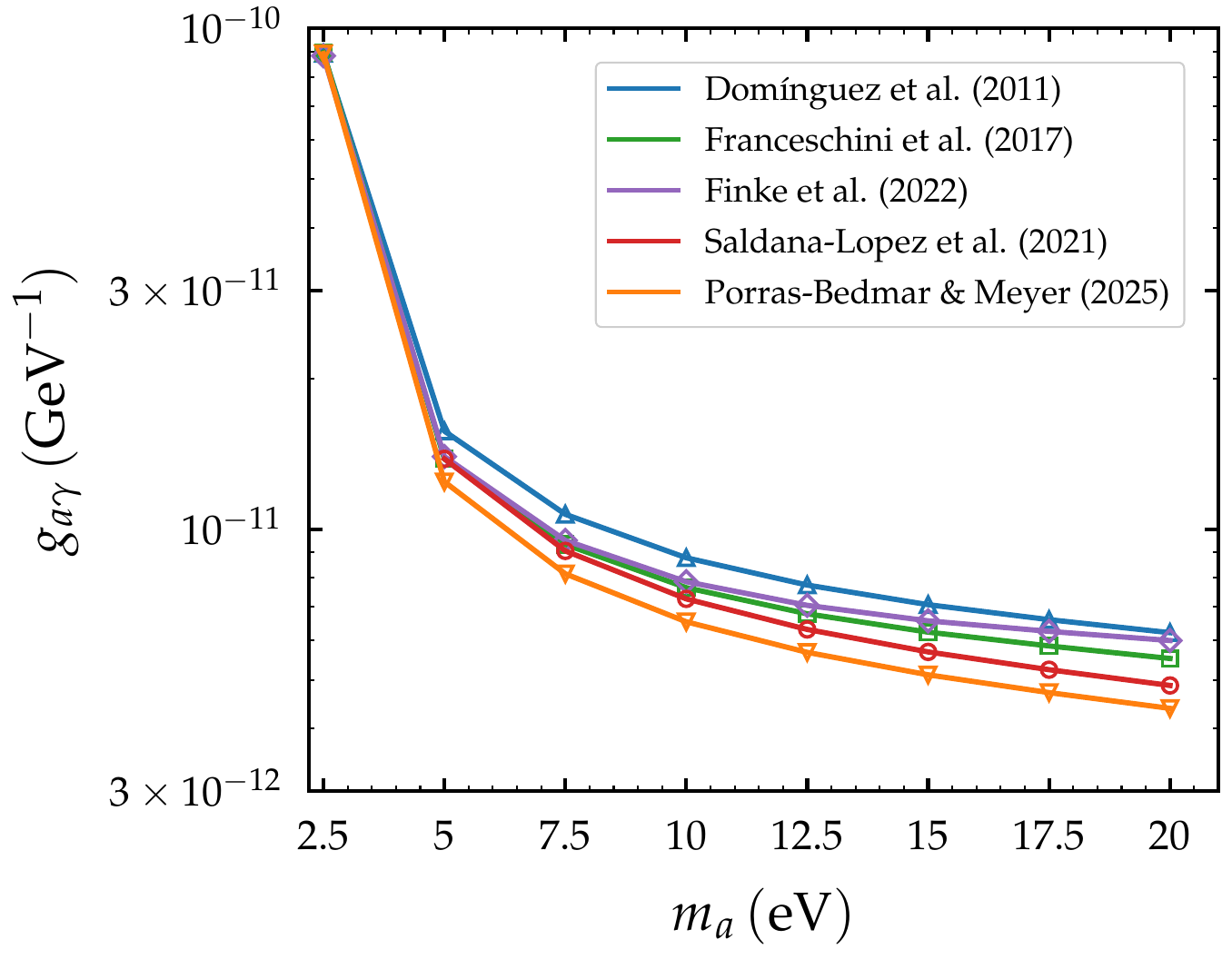}
    \caption{
    {Comparison of the 95\%~C.L.\ upper limits on the photon--ALP coupling $g_{a\gamma}$ as a function of the ALP mass $m_a$, obtained using different EBL models. 
    The curves correspond to the Saldana-Lopez et al.~(2021), Franceschini et al.~(2017), Finke et al.~(2022), Domínguez et al.~(2011), and Porras-Bedmar \& Meyer~(2025) EBL models, as indicated in the legend.}}
    \label{fig:alp_constraints_comparison}
\end{figure}

Among the tested models, we adopt the Domínguez~(2011) EBL model~\cite{2011MNRAS.410.2556D} as our baseline, as it generally yields the most conservative constraints on $g_{a\gamma}$.  
Figure~\ref{fig:alp_constraints_comparison} compares the resulting 95\%~C.L.\ upper limits obtained using the Franceschini et al.~\cite{Franceschini_2017}, Finke et al.~\cite{Finke_2021}, Domínguez et al.~\cite{2011MNRAS.410.2556D}, Saldana-Lopez et al.~\cite{2021MNRAS.507.5144S}, and Porras-Bedmar \& Meyer (using Chary et al. dust templates)~\cite{2025arXiv251020664P} EBL models.  
The overall shapes of the curves are consistent across models, with differences reflecting the relative optical and infrared intensities predicted by each EBL prescription in the wavelength range ($\lambda \sim 0.004$--$4\,\mu$m) relevant for gamma-ray attenuation in the \textit{Fermi}--LAT and H.E.S.S.\ energy bands.  
In particular, the Porras-Bedmar \& Meyer~(2025) model, which employs a different dust emission prescription leading to a slightly brighter optical and near-infrared background, produces the strongest limits, followed by the Saldana-Lopez et al.~(2021) model.  
The Franceschini et al.~(2017) model yields intermediate results, while the Finke et al.~(2022) and Domínguez et al.~(2011) models give comparable limits within uncertainties, with the Domínguez et al.~(2011) model remaining slightly more conservative.

Overall, the variation among models is modest, roughly within a factor of 1.5, confirming that the derived constraints are robust against current uncertainties in EBL modelling.
It should be noted, most of the EBL models considered here are based on galaxy evolution and multiwavelength survey data and do not rely on gamma-ray observations of blazars. An exception is the Finke et al.~(2022) model, which includes gamma-ray constraints; however, the reconstruction remains primarily driven by galaxy emissivities, particularly in the optical band relevant for this analysis. Therefore, potential ALP-induced contributions are not expected to be absorbed into the baseline EBL models.

It is instructive to compare our results with previous works that suggested evidence for heavy ALP decay. Reference~\cite{2020JCAP...03..064K} analysed the combined gamma-ray spectra of VHE blazars using \textit{Fermi}--LAT and IACT data, and reported that a bump-like excess in the EBL spectrum could be produced by ALP decays with $m_a\simeq 2$--3~eV and $g_{a\gamma}\sim 10^{-10}$~GeV$^{-1}$. In Ref.~\cite{2022PhRvL.129w1301B}, the authors investigated the cosmic optical background with \textit{New Horizons}/LORRI and found a $\sim4\sigma$ excess over galaxy counts, which they showed can be explained by ALP dark matter decays with $m_a\simeq 8$--20~eV and $g_{a\gamma}\simeq (3$--$6)\times10^{-11}$~GeV$^{-1}$, assuming $\Omega_a=\Omega_{\rm DM}$.
In contrast, our joint \textit{Fermi}--LAT and H.E.S.S.\ likelihood analysis shows no significant preference for an ALP-induced component. The exclusion contour in Figure~\ref{fig:limits} defines the region disfavoured at 95\%~C.L.  

Our limits reach $g_{a\gamma}\sim7\times10^{-12}$~GeV$^{-1}$ at $m_a\simeq15$~eV, excluding the parameter space suggested in \cite{2022PhRvL.129w1301B} under the assumption $\Omega_a=\Omega_{\rm DM}$. 
It should be noted that for a subdominant ALP component with $\Omega_a < \Omega_{\rm DM}$, the decay-induced EBL contribution scales linearly with $\Omega_a$, implying an additional optical depth proportional to $\Omega_a g_{a\gamma}^2$. In the regime where the constraints are primarily driven by the amplitude of this extra attenuation, the resulting upper limits on $g_{a\gamma}$ are therefore expected to approximately scale as $\Omega_a^{-1/2}$ for fixed $m_a$.
The limits obtained in this study are competitive with existing astrophysical constraints and provide complementary coverage, particularly at higher ALP masses where the decay-induced optical depth modifies the gamma-ray spectra most significantly. The strongest constraints arise near $m_a \sim 20$~eV, consistent with the expected spectral turnover introduced by ALP decay into photons at optical and near-infrared wavelengths.

The parameter region suggested in \cite{2020JCAP...03..064K} around $m_a\simeq2$--3~eV lies at the lower edge of the mass range explored here. In our analysis, we extend the scan down to $m_a = 2.5~\mathrm{eV}$ and do not observe any indication of the corresponding EBL-induced feature in the gamma-ray spectra, and our limits disfavour the parameter region suggested in \cite{2020JCAP...03..064K}.

Within the scanned mass range, our results strengthen existing constraints around
$m_a \sim 15~\mathrm{eV}$ by roughly a factor of two, reaching sensitivities down to
$g_{a\gamma}^{95} \simeq 7\times10^{-12}\,\mathrm{GeV}^{-1}$, which constitute the strongest constraints currently available in this mass range based on gamma-ray observations.
This sensitivity arises because ALP decays in this mass interval produce photons in the optical/UV band, where the EBL most strongly affects TeV gamma-ray propagation.
We note that the analysis is restricted to $m_a \lesssim 20~\mathrm{eV}$, since at higher masses the decay photons lie in the far-UV/soft X-ray regime and no longer contribute significantly to the EBL relevant for gamma-ray attenuation, while also being strongly constrained by existing X-ray observations.

Many of the strongest existing constraints in this mass range (e.g.\ from HST and other spectroscopic observations) are derived from searches for narrow decay lines and therefore depend on assumptions about astrophysical backgrounds and instrumental systematics. 
While these constraints can be very stringent, they probe a different observable and rely on different modelling assumptions than our analysis.

At other masses within the explored range, our limits remain complementary to existing bounds obtained with different techniques.
In particular, while spectroscopic searches for ALP decay lines can provide stronger constraints at some masses, they probe a fundamentally different observable.

Our analysis relies on the integrated decay contribution to the EBL and the resulting modification of the gamma-ray optical depth, and therefore does not depend on individual sources or their D-factors (i.e.\ line-of-sight integrals of the dark matter density relevant for decay signals).
This provides an independent test with different systematics, and therefore an important cross-check of existing constraints, even in regions where existing bounds are stronger. 
In addition, spectroscopic constraints are weakened in parts of the $m_a \sim 10$--$20$~eV range, for example due to astrophysical absorption near the Lyman limit, where our results provide genuinely new constraints.
This makes the constraints derived here robust and complementary across the full mass range $m_a = 2.5$--$20~\mathrm{eV}$.
\section{Conclusions} 
\label{sec:conclusions}

In this paper, we conduct a comprehensive search for ALP decay signatures in the EBL, utilising gamma-ray observations from \textit{Fermi}--LAT and H.E.S.S.\ across a sample of 11 blazars. 
By performing joint spectral fits for each source under both standard and ALP-modified EBL attenuation models, we explore a wide range of ALP masses \(m_a\) and photon--ALP coupling strengths \(g_{a\gamma}\). 
We implement a likelihood-based framework by summing the $C$-statistics across all time-resolved spectral intervals (derived from variability-based segmentation) and constructing the profile likelihood, \(\lambda(m_a, g_{a\gamma})\), to identify disfavoured regions in the \((m_a, g_{a\gamma})\) parameter space.

Our results yield new exclusion limits on ALP parameters, with the strongest constraint obtained at the highest mass tested, \(m_a = 20\,\mathrm{eV}\), where the shorter ALP decay lifetime maximises the decay-induced photon emissivity; at this mass we find \(g_{a\gamma}^{95} \simeq 7 \times 10^{-12}\,\mathrm{GeV}^{-1}\).
In addition, our analysis probes a previously unexplored region of parameter space around \(m_a \sim 10\text{--}20~\mathrm{eV}\), providing new constraints in a mass range where existing bounds are comparatively weaker.
The derived $95\%$~C.L. exclusion contour complements existing astrophysical and laboratory limits, providing independent constraints on ALP properties using high-energy gamma-ray data alone. 
We have also verified that the derived limits are robust against current uncertainties in EBL modelling, with different EBL templates producing only modest variations in the resulting constraints.

This work illustrates the potential of multi-instrument gamma-ray analyses for probing beyond-the-Standard-Model physics. 
Future observations with the Cherenkov Telescope Array Observatory (CTAO; \cite{2019scta.book.....C, 2021JCAP...02..048A}), which will offer increased sensitivity, broader energy coverage, and improved energy and angular resolution, are expected to significantly enhance such searches. 
Extending the present analysis to include multi-TeV observations from wide-field instruments such as the High-Altitude Water Cherenkov Observatory (HAWC; \citep{2024ApJ...972..144A}) or the Large High Altitude Air Shower Observatory (LHAASO; \citep{2016JPhCS.718e2043V}) would further enhance the sensitivity to additional absorption effects induced by the decay of heavy ALPs.
Robust constraints on intrinsic source spectra, however, require complementary coverage at lower energies, provided by instruments such as \textit{Fermi}-LAT and future MeV gamma-ray missions such as AMEGO-X~\citep{2022JATIS...8d4003C}.
A dedicated multi-instrument likelihood analysis represents a natural extension of this work.
In parallel, laboratory experiments and upcoming astrophysical surveys will continue to refine the viable ALP parameter space. 
Joint efforts across observational domains remain essential for uncovering—or constraining—the role of ALPs in cosmology and fundamental physics.

\acknowledgments

The support of the Namibian authorities and of the University of Namibia in facilitating the construction and operation of H.E.S.S. is gratefully acknowledged, as is the support by the Germany Ministry of Research, Technology and Space (BMFTR), the Max Planck Society, the Helmholtz Association, the French Ministry of Higher Education, Research and Innovation, the Centre National de la Recherche Scientifique (CNRS/IN2P3 and CNRS/INSU), the Commissariat à l’énergie atomique et aux énergies alternatives (CEA), the U.K. Science and Technology Facilities Council (STFC), the Irish Research Council (IRC) and the Science Foundation Ireland (SFI), the Polish Ministry of Education and Science, agreement no. 2021/WK/06, the South African Department of Science and Innovation and National Research Foundation, the University of Namibia, the National Commission on Research, Science $\&$ Technology of Namibia (NCRST), the Austrian Federal Ministry of Education, Science and Research and the Austrian Science Fund (FWF), the Australian Research Council (ARC), the Japan Society for the Promotion of Science, the University of Amsterdam and the Science Committee of Armenia grant 21AG-1C085. We appreciate the excellent work of the technical support staff in Berlin, Zeuthen, Heidelberg, Palaiseau, Paris, Saclay, Tübingen and in Namibia in the construction and operation of the equipment. This work benefited from services provided by the H.E.S.S. Virtual Organisation, supported by the national resource providers of the EGI Federation.

The \textit{Fermi}-LAT Collaboration acknowledges generous ongoing support from a number of agencies and institutes that have supported both the development and the operation of the LAT as well as scientific data analysis. These include the National Aeronautics and Space Administration and the Department of Energy in the United States, the Commissariat à l'Energie Atomique and the Centre National de la Recherche Scientifique / Institut National de Physique Nucléaire et de Physique des Particules in France, the Agenzia Spaziale Italiana and the Istituto Nazionale di Fisica Nucleare in Italy, the Ministry of Education, Culture, Sports, Science and Technology (MEXT), High Energy Accelerator Research Organization (KEK) and Japan Aerospace Exploration Agency (JAXA) in Japan, and the K. A. Wallenberg Foundation, the Swedish Research Council and the Swedish National Space Board in Sweden.
Additional support for science analysis during the operations phase is gratefully acknowledged from the Istituto Nazionale di Astrofisica in Italy and the Centre National d'Etudes Spatiales in France. This work performed in part under DOE Contract DE- AC02-76SF00515.

The authors would like to thank Alberto Domínguez and Pedro De la Torre Luque for their helpful comments and suggestions during the internal review process.

\clearpage
\appendix
\section{Appendix}
\label{Appendix}


Figures~\ref{fig:lc_A1a}--\ref{fig:lc_A1c} show the nightly-binned light curves and the corresponding
Bayesian-block segmentation for all sources in the sample.


\setcounter{figure}{0}
\renewcommand{\thefigure}{A1\alph{figure}}

\begin{figure} [!b]
    \centering
    \includegraphics[width=\linewidth]{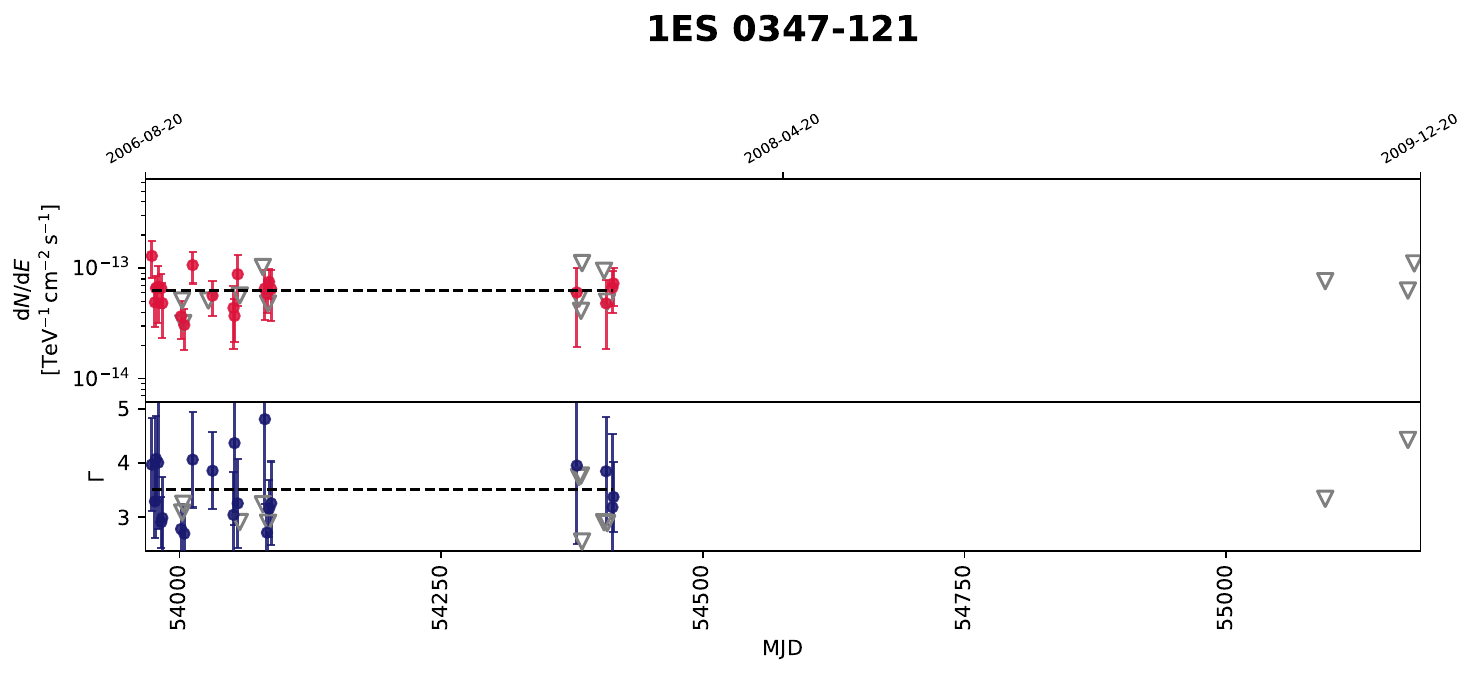}
    \includegraphics[width=\linewidth]{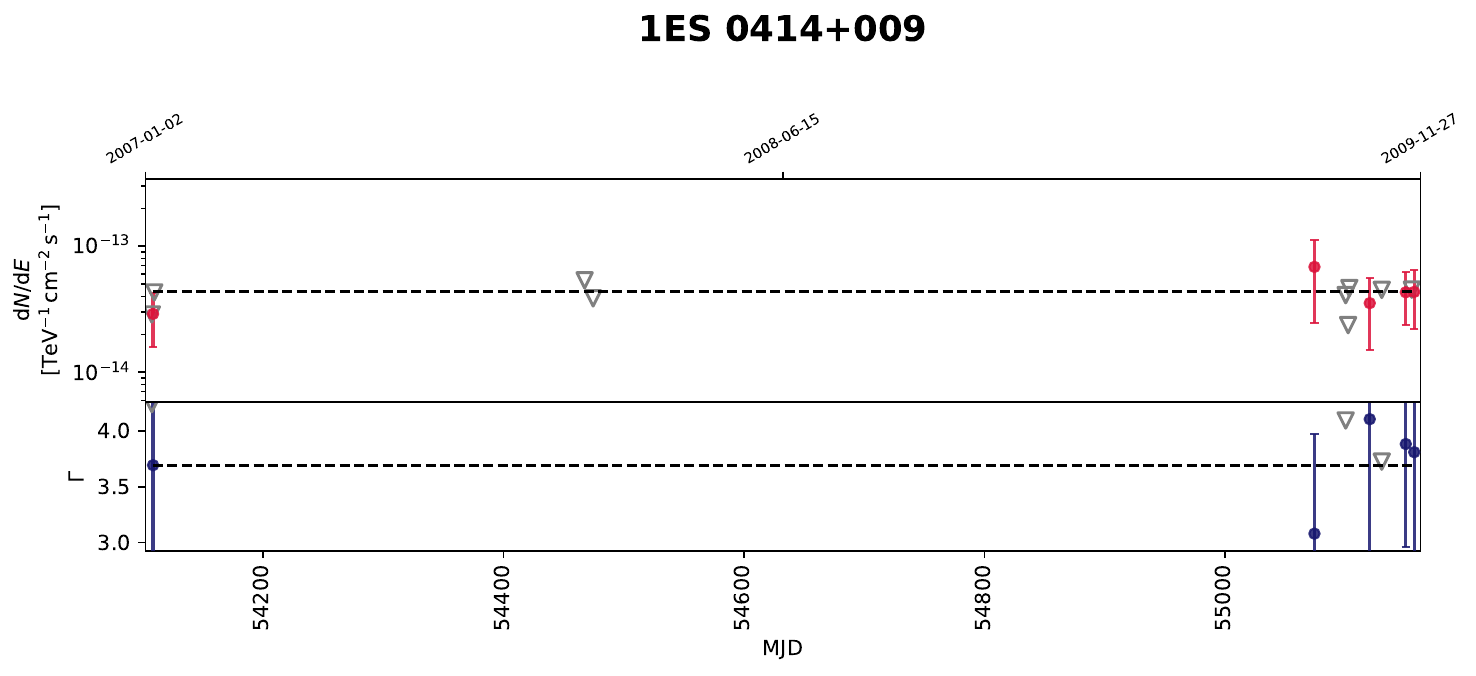}

    \caption{
H.E.S.S.\ nightly-binned light curves over the analysed time intervals.
For each source, the top panel shows the flux evolution, with detections shown in red and 95\% upper limits
indicated by grey inverted triangles. Horizontal dashed lines mark the mean flux levels within the Bayesian-block
segments. The bottom panel shows the evolution of the photon index~$\Gamma$ derived from the nightly spectra, with
dashed lines indicating the corresponding Bayesian-block segmentation of the index. The analysed time intervals
are listed in Table~\ref{tab:combined_blazar_summary}.
}
    \label{fig:lc_A1a}
\end{figure}

\begin{figure}[!p]
    \centering
    \includegraphics[width=\linewidth]{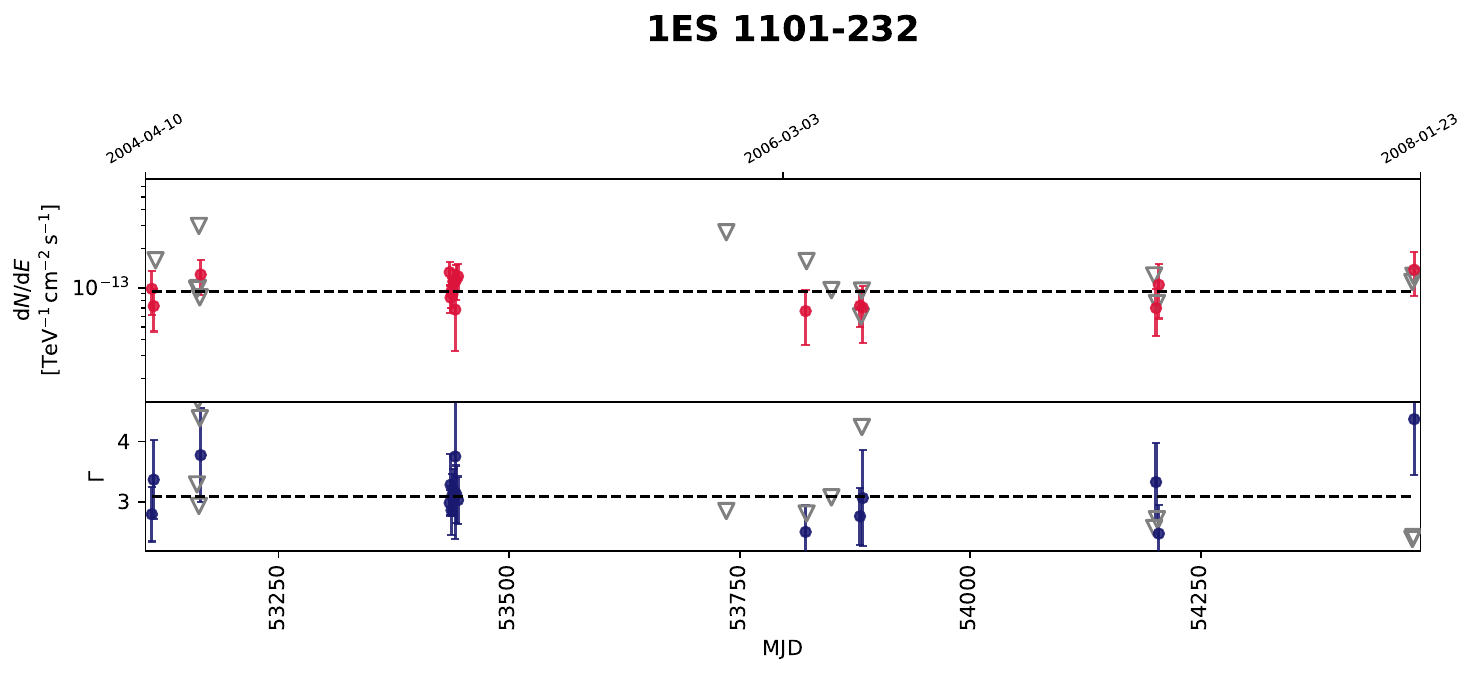}
    \includegraphics[width=\linewidth]{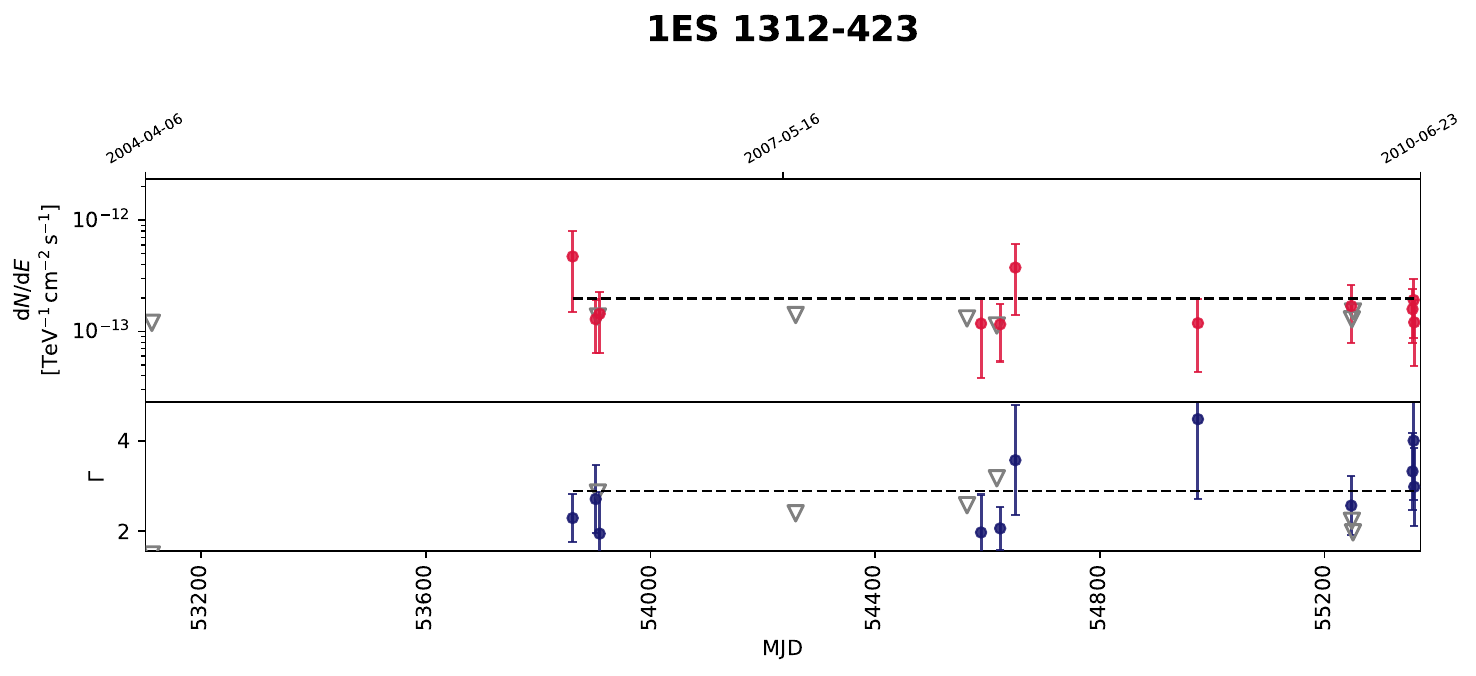}
    \includegraphics[width=\linewidth]{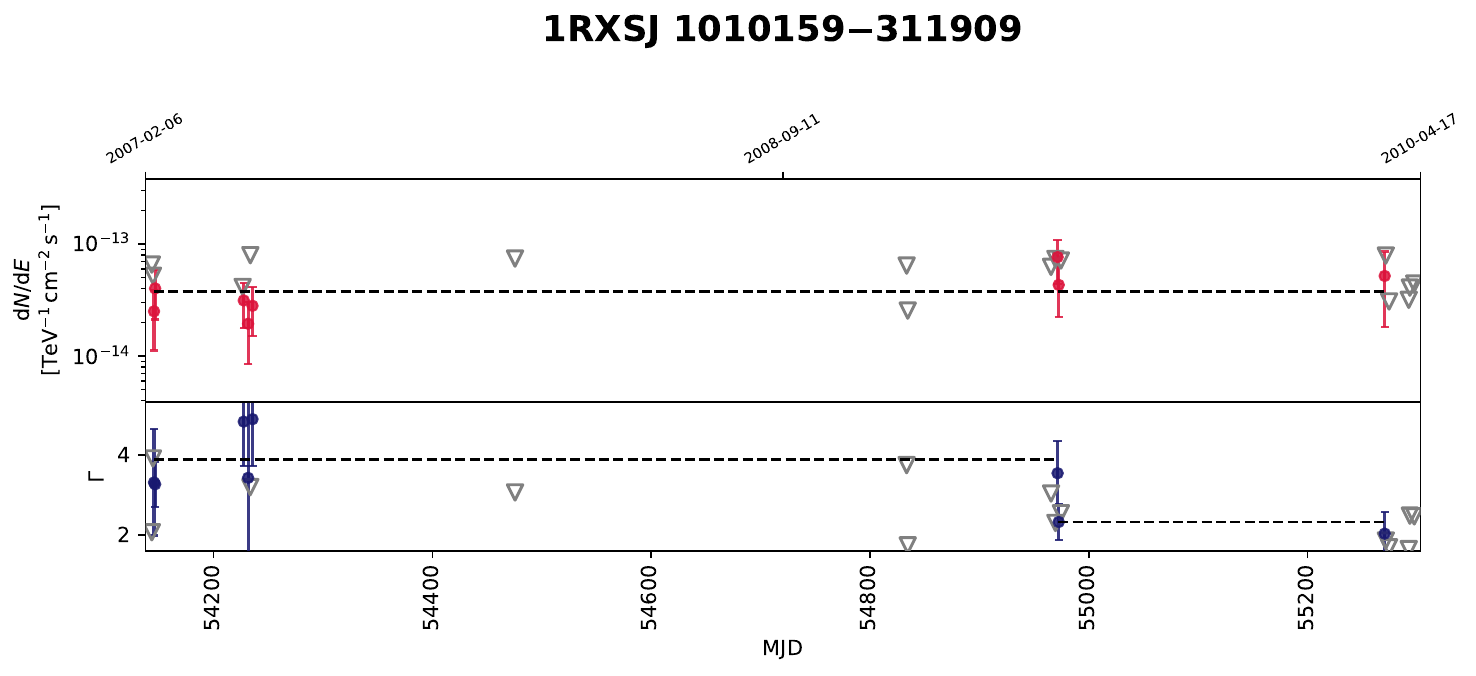} 
    
    \caption{Same as Fig.~\ref{fig:lc_A1a}, continued.}
    \label{fig:lc_A1b}
\end{figure}

\begin{figure}[!p]
    \centering
    \includegraphics[width=\linewidth]{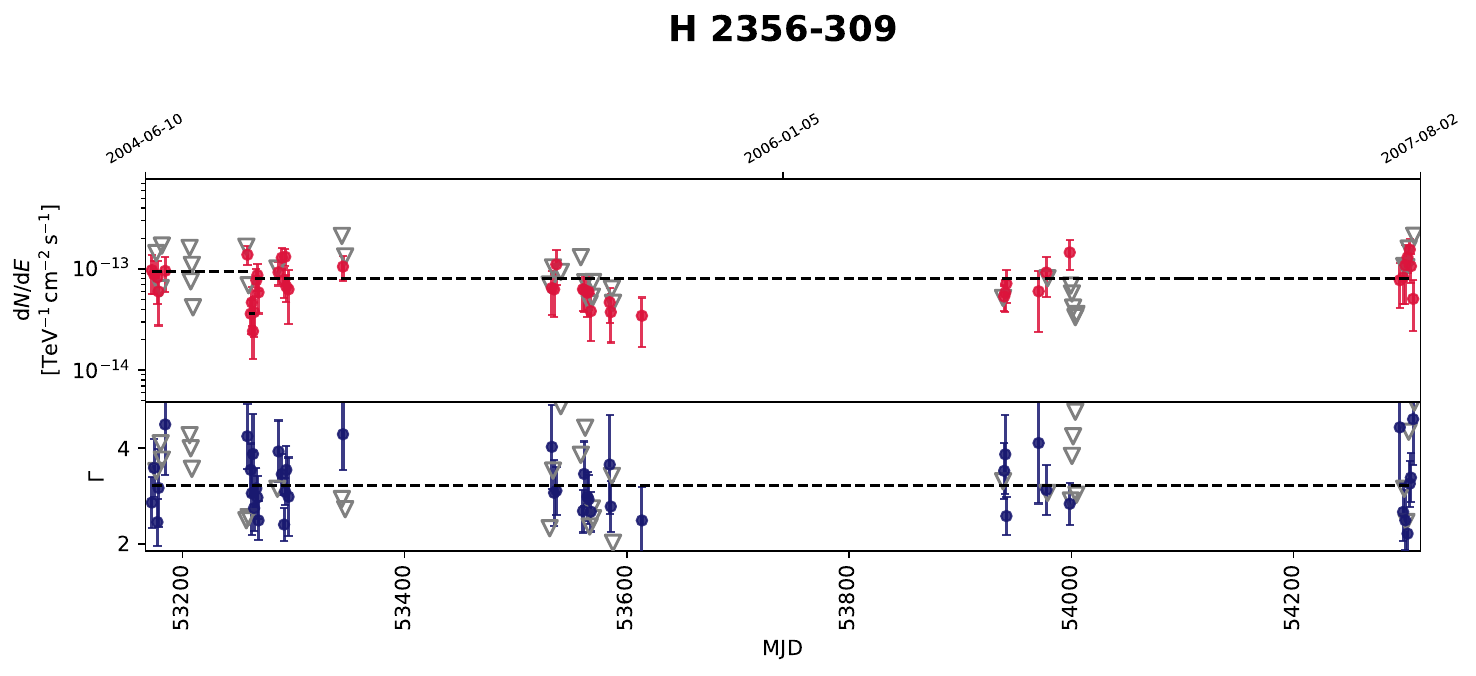}    
    \includegraphics[width=\linewidth]{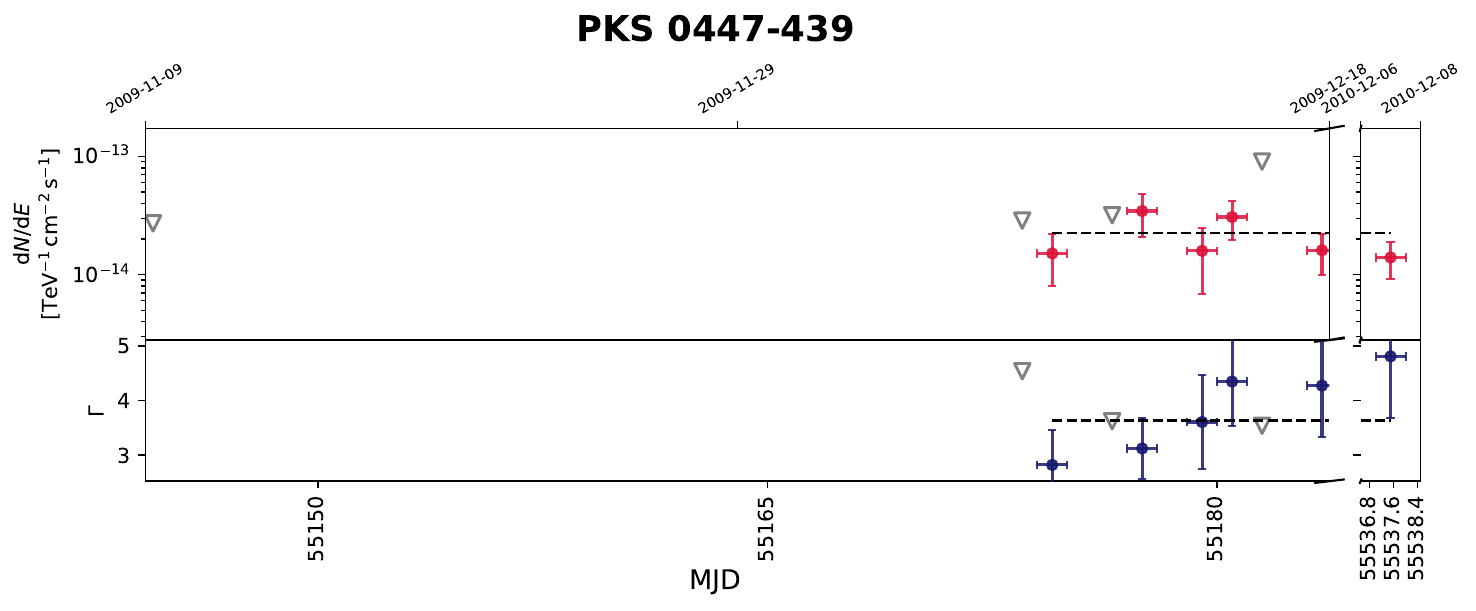}
    \includegraphics[width=\linewidth]{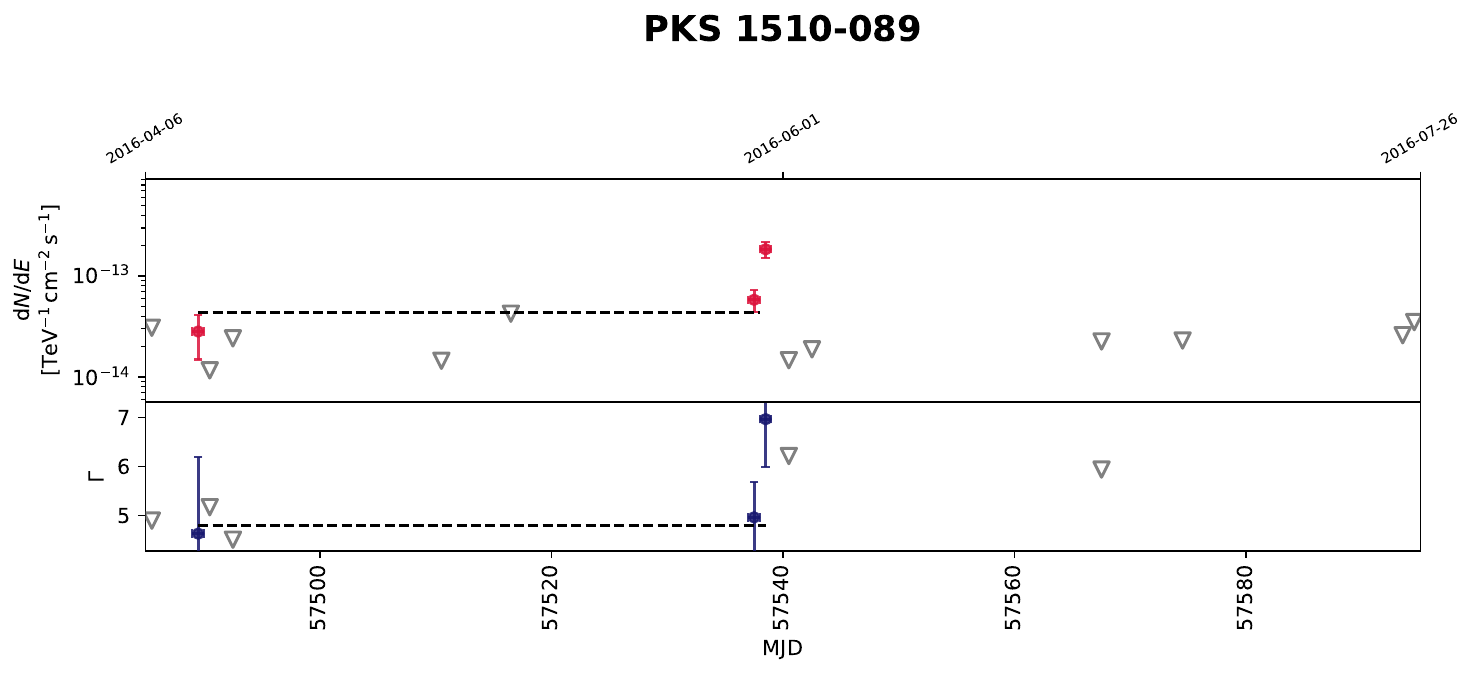}

    \caption{Same as Fig.~\ref{fig:lc_A1a}, continued.}
    \label{fig:lc_A1c}
\end{figure}

\FloatBarrier

\renewcommand{\thefigure}{A\arabic{figure}}
\setcounter{figure}{1} 

\begin{figure}[H]
    \centering
    \includegraphics[width=0.5\textwidth]{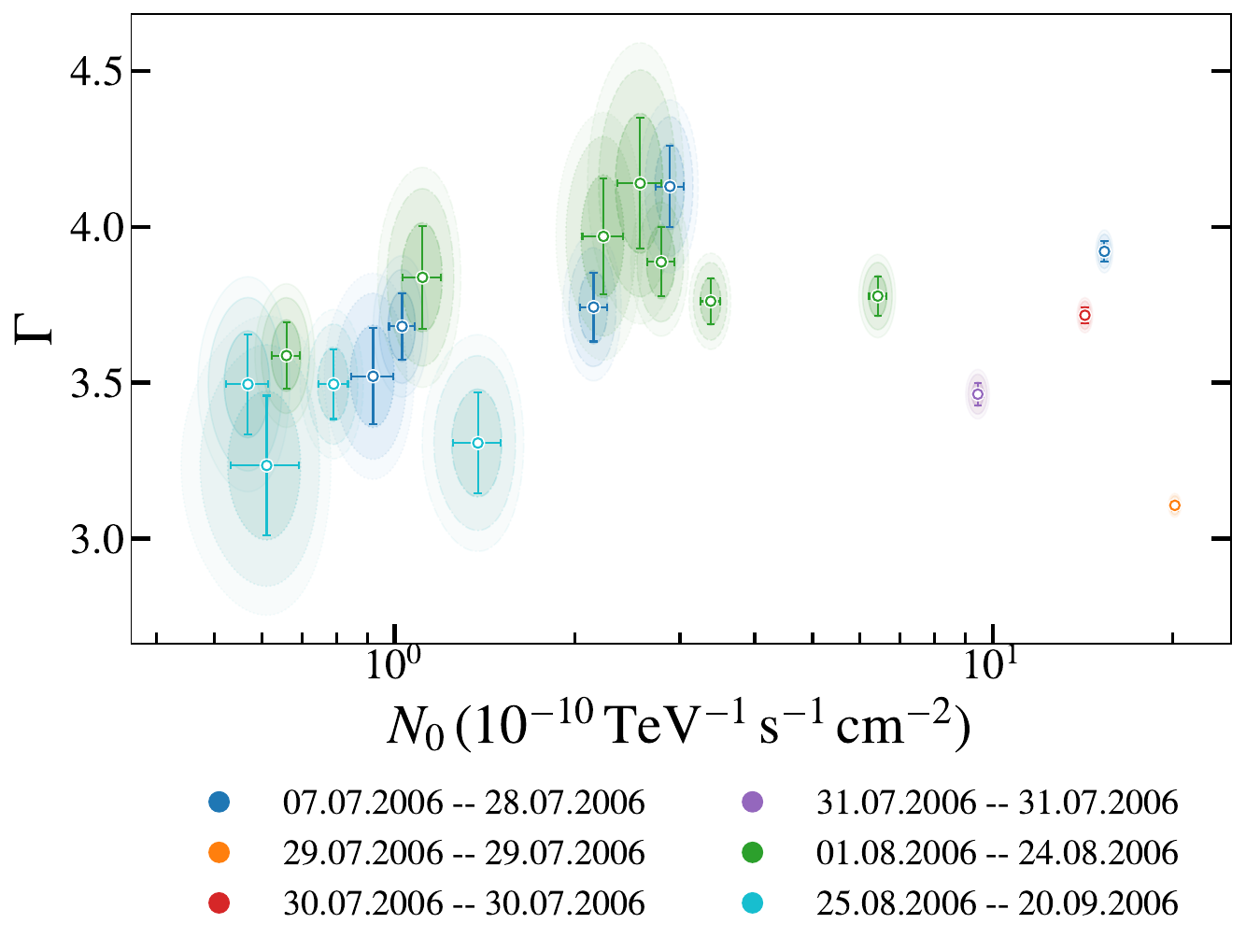}
    \caption{
Best-fit photon index $\Gamma$ versus flux normalisation $N_0$ for PKS~2155--304, shown with $1\sigma$, $2\sigma$,
and $3\sigma$ confidence ellipses for each Bayesian-block interval identified in the 2006 H.E.S.S.\ light curve.
}
    \label{fig:bestfit_ellipses_pks2155}
\end{figure}

Figure~\ref{fig:bestfit_ellipses_pks2155} illustrates the segmentation procedure for PKS~2155--304 by showing the
best-fit photon index $\Gamma$ versus flux normalisation $N_0$ for each Bayesian-block interval. Blocks that exhibit
overlapping confidence ellipses are consistent with a single spectral state, while those showing significant separation
motivate the segmentation adopted in our analysis. In practice, this criterion is applied iteratively, such that blocks can be merged if they are connected through pairwise overlaps, even if not all confidence ellipses overlap simultaneously.

\vspace{-0.5em}

\begin{figure}
  \centering
  \begin{subfigure}{0.4\textwidth}
    \includegraphics[width=\linewidth]{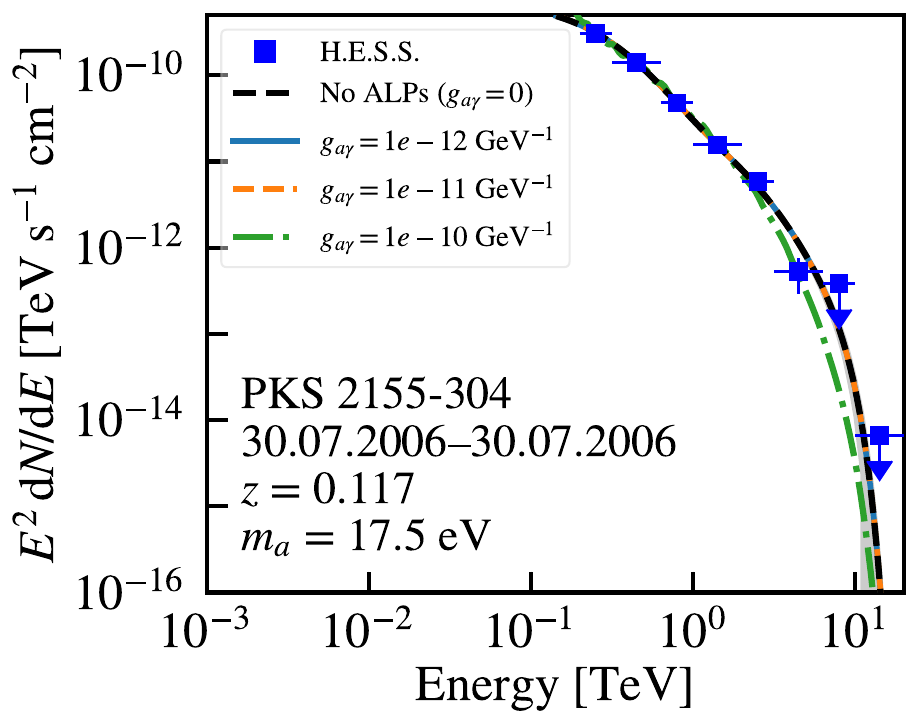}
    \label{fig:pks2155_a}
  \end{subfigure}\hfill
  \begin{subfigure}{0.4\textwidth}
    \includegraphics[width=\linewidth]{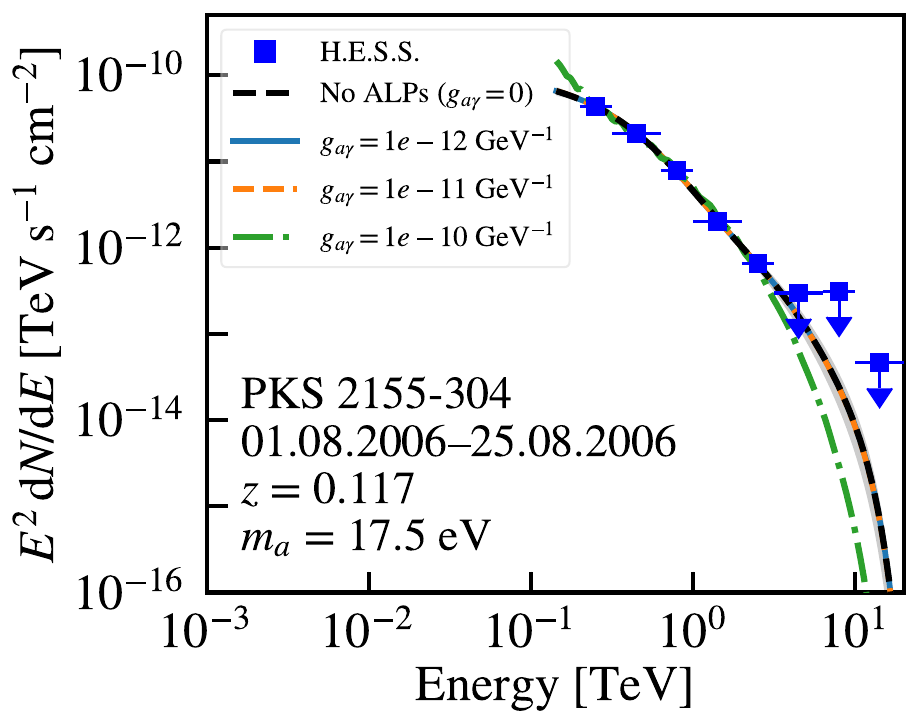}
    \label{fig:pks2155_b}
  \end{subfigure}
\caption{
Spectral energy distributions (SEDs) of PKS~2155--304 for the two flare episodes corresponding to the strongest local preference for an ALP contribution in our sample.
H.E.S.S.\ flux points (squares) are shown together with the best--fit no-ALP scenario and ALP
models for representative photon--axion couplings
(\(g_{a\gamma}=10^{-12},\,10^{-11},\,10^{-10}\ \mathrm{GeV}^{-1}\)).
In both panels, the ALP mass is fixed to \(m_a = 17.5\,\mathrm{eV}\). The best-fit spectral models overlap for the no-ALP scenario and the two lowest photon--axion couplings
(\(g_{a\gamma}=10^{-12},\,10^{-11}\ \mathrm{GeV}^{-1}\)).
}

  \label{fig:pks2155_seds}
\end{figure}

Figure~\ref{fig:pks2155_seds} shows the spectral energy distributions for the two PKS~2155--304 segments that contribute
most strongly to the combined profile likelihood. These intervals correspond to the highest values of
$\lambda(m_a, g_{a\gamma}=0)$ and therefore to the largest local preference for an ALP-induced optical-depth component.

\FloatBarrier


\clearpage
\bibliographystyle{jhep}  
\bibliography{draft_paper}



\end{document}